\documentclass[a4paper, onecolumn, 10pt, accepted=2025-12-01, dvipsnames]{quantumarticle}
\pdfoutput=1

\usepackage[utf8]{inputenc}
\usepackage[english]{babel}
\usepackage[T1]{fontenc}
\usepackage{amsmath}

\usepackage{hyperref}

\usepackage[dvipsnames, table]{xcolor}
\definecolor{quantumviolet}{HTML}{53257F}
\definecolor{quantumgray}{HTML}{555555}

\usepackage[numbers,sort&compress]{natbib}

\usepackage{bm}
\usepackage{bbm}
\usepackage[retainorgcmds]{IEEEtrantools}
\usepackage{graphicx}
\usepackage{mathrsfs}
\usepackage{amsfonts}
\usepackage{amssymb}
\usepackage{color}
\usepackage{times,txfonts}
\usepackage{nicefrac}
\usepackage{ragged2e} 
\usepackage{physics} 
\usepackage{verbatim} 
\usepackage{float}
\usepackage{blindtext}
\usepackage{lipsum}  
\usepackage[inkscapearea=page]{svg} 
\usepackage[export]{adjustbox}

\usepackage{multirow} 
\usepackage{makecell} 

\usepackage{algorithm}
\usepackage[noend]{algpseudocode}

\definecolor{cadmiumgreen}{HTML}{097969}

\newcommand{\soliton}{\includegraphics[valign=c]{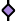}}
\newcommand{\leg}{\includegraphics[valign=c]{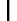}}
\newcommand{\cphasedraw}{\includegraphics[valign=c]{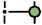}}
\newcommand{\swapdraw}{\includegraphics[valign=c]{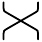}}

\newcommand{\expec}[1]{\langle #1 \rangle}
\newcommand{\swap}{\mathrm{SWAP}}

\newcommand{\id}{\mathbbm{1}}
\newcommand{\phasegate}{\phi}
\newcommand{\CPhase}{\mathrm{CP}(\phasegate)}
\newcommand{\scatteringphase}{\Phi^{(m)}}

\newcommand\source[1]{%
    \tikz[remember picture,baseline,inner sep=0pt] {%
        \node [name=source,anchor=base]{$#1$};
    }%
    \setcounter{target}{0}
}

\newcounter{target}
\newcommand\target[1]{%
    \tikz[remember picture,baseline,inner sep=0pt] {%
        \node [name=target-\thetarget, anchor=base]{$#1$};
    }%
    \stepcounter{target}%
}

\newcommand\drawarrows{
    \tikz[remember picture, overlay, bend left=20, -latex] {
        \foreach \i [evaluate=\i as \n using int(\i-1)] in {1,...,\thetarget} {
            \draw [<->, thick] ([yshift=3pt]source.north) to ([yshift=3pt]target-\n.north);
        }
    }
}

\begin{document}

\title{Probes of Full Eigenstate Thermalization in Ergodicity-Breaking Quantum Circuits}

\author{Gabriel O. Alves}
\email{alves.go.co@gmail.com}
\affiliation{Max Planck Institute for the Physics of Complex Systems, 01187 Dresden, Germany}

\author{Felix Fritzsch}
\affiliation{Max Planck Institute for the Physics of Complex Systems, 01187 Dresden, Germany}

\author{Pieter W. Claeys}
\affiliation{Max Planck Institute for the Physics of Complex Systems, 01187 Dresden, Germany}

\begin{abstract}
   The eigenstate thermalization hypothesis (ETH) is the leading interpretation in our current understanding of quantum thermalization.  
   Recent results uncovered strong connections between quantum correlations in thermalizing systems and the structure of free probability theory, leading to the notion of full ETH.
   However, most studies have been performed for ergodic systems and it is still unclear whether or how full ETH manifests in ergodicity-breaking models.
   We fill this gap by studying standard probes of full ETH in ergodicity-breaking quantum circuits, presenting numerical and analytical results for interacting integrable systems.
   These probes can display distinct behavior and undergo a different scaling than the ones observed in ergodic systems. 
   For the analytical results we consider an interacting integrable dual-unitary model and present the exact eigenstates, allowing us to analytically express common probes for full ETH.
   We discuss the underlying mechanisms responsible for these differences and show how the presence of solitons dictates the behavior of ETH-related quantities in the dual-unitary model.
   We show numerical evidence that this behavior is sufficiently generic away from dual-unitarity when restricted to the appropriate symmetry sectors.
\end{abstract}

\maketitle{}


\section{Introduction}

A great triumph of classical thermodynamics, be it through macroscopic or microscopic descriptions, lies in its ability to satisfactorily provide meaningful insights about nature in spite of the intricate interactions and extensive degrees of freedom found in real physical systems.   
Had thermodynamics not emerged as a theory, reaching similar results solely through classical mechanics would certainly become an insurmountable challenge \cite{callenThermodynamicsIntroductionThermostatistics1985}.
Such motivation is not limited to classical systems and, just as one would never contemplate describing a classical gas in terms of its full configuration space, it would be natural to further extend this perspective to quantum mechanics.
Making general statements about quantum many-body systems and their features, regardless of their complexity and irrespective of the particular details therein, becomes then a central task \cite{tasakiQuantumDynamicsCanonical1998}.
However, because of the idiosyncrasies present in both fields, extending thermodynamics to the realm of quantum mechanics comes with its own set of obstacles~\cite{Nazir2018}. 

\begin{figure}[t!]
    \centering
    \includegraphics[width=0.7\columnwidth]{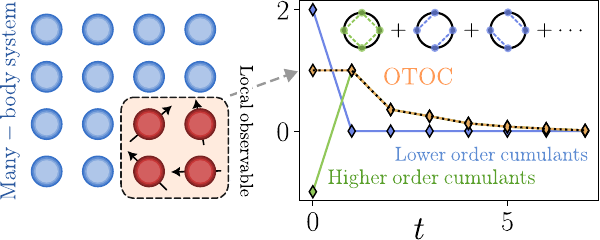}
    \caption{According to the eigenstate thermalization hypothesis (ETH), eigenstates of quantum many-body systems appear thermal whenever local observables are probed. 
    For such ergodic systems, the dynamics of out-of-time order correlators can be decomposed in the sum of thermal cumulants -- an ETH-based analogue of free cumulants.
    This decomposition is based on the combinatorics of non-crossing partitions and correctly captures the long-time behavior of OTOCs.}
    \label{fig:eth_illustration}
\end{figure}

One of the leading formulations to tackle this type of problem is the eigenstate thermalization hypothesis (ETH), which states that eigenstates of quantum many-body  systems look thermal whenever local observables are considered~\cite{deutsch1991quantum,
srednickiChaosQuantumThermalization1994b,
dalessioQuantumChaosEigenstate2016a, 
deutschEigenstateThermalizationHypothesis2018}. 
More concretely, the diagonal components of such observables agree with the microcanonical prediction from statistical mechanics. 
Meanwhile, their off-diagonal part encodes information about fluctuations and is described by random matrix behavior, with some additional structure embedded on spectral functions, reflecting the physical features of the model.

An important and recent advance in the field is the concept of \emph{full ETH}, an extension of the existing ETH ansatz designed to encompass higher-order correlations, as initially introduced by L. Foini and J. Kurchan \cite{foiniEigenstateThermalizationHypothesis2019a}. 
Subsequently, S. Pappalardi and both authors unveiled a relationship between full ETH and free probability (FP) theory: a link which lies in the combinatoric nature of their structure, described in terms of non-crossing partitions~\cite{pappalardiEigenstateThermalizationHypothesis2022}.
Free probability was originally kick-started by D. Voiculescu, first appearing in the 1980s in the study of operator algebras in a mathematical context \cite{voiculescuSymmetriesReducedFree1985, voiculescuAdditionCertainNoncommuting1986a, voiculescuMultiplicationCertainNonCommuting1987}. 
Since then, the field has seen growing interest and an increasing range of applications \cite{novakThreeLecturesFree2012, mingoFreeProbabilityRandom2017}, with full ETH being one of them.
In the latter context, the formalism of free probability has allowed to characterize contributions to higher-order correlation functions predicted by the full ETH ansatz in ergodic systems
~\cite{foiniEigenstateThermalizationHypothesis2019a,
pappalardiGeneralEigenstateThermalization2025,  
pappalardiMicrocanonicalWindowsQuantum2024, 
jindalGeneralizedFreeCumulants2024,
fritzschMicrocanonicalFreeCumulants2025}. 
Moreover, FP suggests that under chaotic evolution time-evolved and static observables become `freely' independent~\cite{valliniLongtimeFreenessKicked2024, 
chenFreeIndependenceNoncrossing2024a,
camargoQuantumSignaturesChaos2025}.
Free independence was then also linked to the emergence of unitary designs from chaotic evolution~ \cite{favaDesignsFreeProbability2023a}.

Applying concepts from FP allows for making predictions for $n$-point correlation functions: These can be decomposed in recursively defined thermal free cumulants, for which full ETH provides an explicit description in terms of specific summations over matrix elements. 
Such free cumulants can also be defined for nonergodic systems, but they cannot be expected to exhibit thermal behavior due to the additional conservation laws in integrable systems.
However, it is still an open question if the overall description of free cumulants provided by full ETH applies in nonergodic systems.
We here answer this question in the affirmative for a class of interacting integrable quantum circuits by studying aspects of full ETH and FP in ergodicity-breaking circuits, with a particular focus on dual-unitary models.
Dual-unitary (DU) circuits consist in a special class of quantum circuit models whose gates display the special feature of being unitary in both temporal and spatial directions~\cite{gopalakrishnan_unitary_2019,bertiniExactCorrelationFunctions2019}.
Although generic many-body models are, even numerically, often exponentially hard to solve, this richer structure of DU circuits makes their dynamics analytically treatable even for chaotic models.
Because of such properties, DU circuits have emerged as a valuable minimal model of quantum many-body dynamics, making them a stepping stone to understand more complex phenomena in quantum information and condensed matter.
Nonergodic DU circuits have recently gained increasing interest since the combination of nonergodicity and dual-unitarity strongly constrains the dynamics and allows for models that support solitons~\cite{
bertini_operator_2020,
gombor_superintegrable_2022,
holden-dyeFundamentalChargesDualunitary2023,
sommersCrystallineQuantumCircuits2023,
claeysOperatorDynamicsEntanglement2024,
folignoNonequilibriumDynamicsCharged2025}. 
While interacting, these support an analytical treatment of the dynamics of e.g. local operator entanglement and scrambling~\cite{bertini_operator_2020,dowling2023scrambling}, information recovery protocols~\cite{rampp_hayden-preskill_2024}, temporal entanglement~\cite{giudice_temporal_2022} and magic spreading~\cite{montana_lopez_exact_2024}, all tractable 
using either standard dual-unitary techniques or Bethe ansatz methods in generic interacting integrable systems.

In this paper we demonstrate that central predictions of full ETH are obeyed by interacting integrable circuits for local observables, once one restricts the discussion to the symmetry sectors corresponding to the relevant conservation laws.
In particular, we consider the statistics of off-diagonal matrix elements.
We show that their fourth moment, associated to a so-called crossing partition in the framework of FP, is suppressed with the dimension of the symmetry sector, as predicted by full ETH (as reviewed in Sec.~\ref{sec:FP_ETH}). 
Correlations between off-diagonal matrix elements at different (quasi-)energies, associated with non-crossing partitions in FP, are additionally shown to factorize into lower order correlations, again consistent with the full ETH ansatz.
We first obtain those results within interacting integrable dual-unitary circuits by analytically obtaining the eigenstates and explicitly relating the resulting statistics of matrix elements to conservation laws~ (Sec.~\ref{sec:DU_circuits}).
By obtaining the analytical eigenstates of the model we are then able to understand the underlying mechanisms explaining the statistics of off-diagonal elements and the full ETH probes in the model.
These results imply that the dynamics within the symmetry sectors, while non-mixing, can be treated as ergodic.
We subsequently provide a heuristic argument for the fate of our results away from dual-unitary models, which we confirm numerically (Sec.~\ref{sec:nondu}).
More precisely, we find the contributions from crossing partitions to be suppressed with the dimension of the charge sector for the most local conservation law, e.g. magnetization.
This scaling gives rise to polynomial corrections on the exponential suppression with system size and resembles what is to be expected in chaotic models with a single local conservation law. 
Breaking integrability, we indeed recover the results expected in such models (Sec.~\ref{sec:nonint}).
We conclude with a discussion and outlook (Sec.~\ref{sec:conclusion}).


\section{Free Probability Theory and full ETH}\label{sec:FP_ETH}

The eigenstate thermalization hypothesis (ETH) presents an ansatz for the matrix elements of a local observable $A$ in the eigenbasis of a model with eigenenergies $E_i$ and corresponding eigenstates $\ket{E_i}$ \cite{dalessioQuantumChaosEigenstate2016a}. 
The model and the observable are said to satisfy ETH if they obey the following ansatz:
\begin{equation}\label{eq:standard_ETH}
    A_{ij} \equiv \langle E_i | A | E_j \rangle
    =
    \bar{A}(E)\, \delta_{ij}
    + e^{-S(E^+_{ij})}f_{E^+_{ij}}(\omega_{ij}) R_{ij},
\end{equation}
where $R_{ij}$ is a (possibly complex) Gaussian distributed random variable with mean zero and unit variance. 
Similarly, one also defines the average energy $E \equiv (E_i + E_j)/2$ and the energy difference $\omega_{ij} \equiv E_i - E_j$. Finally, $\bar{A}(E)$ is the microcanonical expectation of $A$ and corresponds to the so-called {diagonal} ETH, whereas $S(E)$ is the thermodynamic entropy and $f_E(\omega)$ is the (second-order) spectral function. These latter quantities are associated with the {off-diagonal} ETH.
The first term in the ansatz gives rise to the agreement between the late time steady state of the observable and the statistical mechanics prediction, while the second term guarantees fluctuations
around the latter to be small~~\cite{deutsch1991quantum,
srednickiChaosQuantumThermalization1994b,
dalessioQuantumChaosEigenstate2016a, 
deutschEigenstateThermalizationHypothesis2018}. 
This ansatz bridges quantum dynamics with statistical physics, providing a phenomenology as to why many generic quantum many-body models appear to thermalize locally, despite undergoing a global unitary evolution. 

However, the ETH ansatz above is not able to capture higher-order correlations, i.e. general $n$-point functions.
Such higher-order correlations can be expressed in higher moments of off-diagonal matrix elements, which for Gaussian variables follow from the variance and hence from the two-point function.
That is, all higher-order correlators would be determined by at most two-point functions, which is not true in general
~\cite{pappalardiGeneralEigenstateThermalization2025, 
chanEigenstateCorrelationsThermalization2019, 
hahnEigenstateCorrelationsEigenstate2024}
-- additional information is gained by studying higher-point functions.
The so-called \emph{full} eigenstate thermalization hypothesis was introduced in Ref.~\cite{foiniEigenstateThermalizationHypothesis2019a} to address this issue. 
To motivate full ETH, note that an equivalent way of expressing Eq.~\eqref{eq:standard_ETH} is in terms of the average, e.g., over small energy windows, of these different matrix elements.
For the diagonal and off-diagonal elements this yields
\begin{equation}\label{eq:standard_eth_means}
    \overline{A_{ii}} = \bar{A}(E), \quad
    \overline{A_{ij} A_{ji}} = e^{-S(E^+_{ij})}f_{E^+_{ij}}(\omega_{ij}) ,
\end{equation}
since $\overline{R_{ij}} = 0$ and $\overline{|R_{ij}|^2} = 1$.
Therefore, in analogy to Eq.~\eqref{eq:standard_eth_means}, it was argued in Ref.~\cite{foiniEigenstateThermalizationHypothesis2019a} that higher order correlations among the matrix elements can be captured using a full eigenstate thermalization hypothesis of the form 
\begin{equation}\label{eq:general_eth}
    \overline{A_{i_1 i_2} A_{i_2 i_3} ... A_{i_q i_1}} 
    = 
    e^{-(q - 1)S(E^+)}
    f_{E^+}^{(q)}(\omega_{i_1 i_2}, ..., \omega_{i_{q-1} i_q}),
\end{equation}
for distinct indices $i_1 \neq i_2 \neq .... \neq i_q$.
Here we have extended the definition of average energy to 
$E^+ \equiv (E_{i_1} + ... + E_{i_q})/q$.
The entropic term sets the magnitude of these correlations, while the smooth, order 1 functions $f_{E^+}^{(q)}(\omega_{i_1 i_2}, ..., \omega_{i_{q-1} i_q})$ capture their energy dependence.
\begin{figure}
    \centering
    \includegraphics[width=\textwidth]{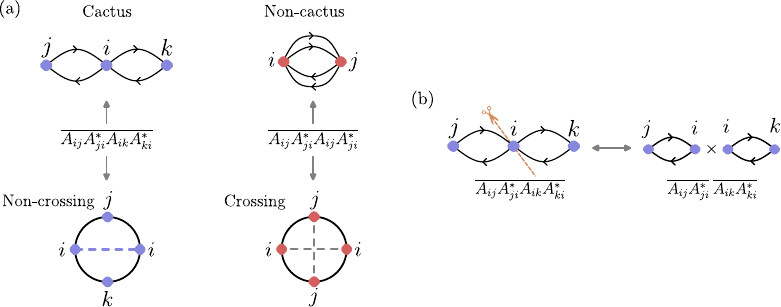}
    \caption{(a) Example of cactus and non-cactus diagrams depicting the generalized ETH averages of Eq.~\eqref{eq:general_eth}.
    The cacti in the ETH picture correspond to non-crossing partitions as understood by free probability theory.
    The correspondence between non-cactus diagrams and crossing partitions is analogous.
    (b) Illustration of the factorization of the cactus diagrams in terms of its `leaves'.
    That is, complicated averages corresponding to cacti can be factorized into simple averages of lower order, as written in Eq.~\eqref{eq:eth_factorization}.
    }
    \label{fig:partition_eth_correspondence}
\end{figure}
The combinatorics underlying the energy indices in Eq.~\eqref{eq:general_eth} can be expressed as a so-called ETH diagram~\cite{pappalardiEigenstateThermalizationHypothesis2022}, as illustrated in Fig.~\ref{fig:partition_eth_correspondence}, closely related to the diagrammatics of free probability~\cite{speicherMultiplicativeFunctionsLattice1994}. 
We can use these diagrams to introduce two central properties of the generalized ansatz in Eq.~\eqref{eq:general_eth}:

\begin{enumerate}
    \item Cactus diagrams are diagrams in which simple loops have at most a single vertex in common. Products associated with cactus diagrams \emph{factorize} in terms of product of its leaves, i.e. into irreducible simple loops:
    \begin{equation}\label{eq:eth_factorization}
        \begin{split}
            \centering
            \overline{A_{i_1 i_2} A_{i_2 i_3} ... A_{i_q i_1}} 
            =
            \overline{A_{i_1 i_2} A_{i_2 i_3} ... A_{i_{k-1} i_1}} 
            \:\,
            \overline{A_{i_1 i_{k + 1}} A_{i_{k+1} i_{k+2}} ... A_{i_q i_1}}.
        \end{split}
    \end{equation}
    \item Non-cactus diagrams are subleading, and are expected to vanish as $1/D$, with $D$ the dimension of the many-body Hilbert space, in the thermodynamic limit, e.g.:
    \begin{equation}\label{eq:eth_crossing}
    \frac{1}{D}
    \sum_{i \neq j}
    \overline{|A_{ij}|^2}\:\overline{|A_{ij}|^2}
    \sim
    O\left( \frac{1}{D}\right).
    \end{equation}
\end{enumerate}

These properties are closely related to so-called \emph{free cumulants} in FP (see Fig.~\ref{fig:partition_eth_correspondence}). 
Free cumulants extend the notion of cumulants for commuting variables to noncommuting variables, and can be indexed by a so-called non-crossing partition~\cite{speicherMultiplicativeFunctionsLattice1994}. A defining property in the structure of cumulants in FP is that -- in contrast to standard probability theory -- crossing partitions vanish in the computation of statistical cumulants, and free cumulants factorize according to the blocks in their associated non-crossing partition. 
The cactus diagrams can be associated with the non-crossing partitions, while the non-cactus diagrams can be associated with the (vanishing) crossing partitions.
This motivates the notion of \emph{thermal free cumulants}, which are implicitly defined in terms of $n$-point correlation functions~\cite{pappalardiEigenstateThermalizationHypothesis2022}:
\begin{equation}\label{eq:thermal_cumulants}
    \langle A(t_1) ... A(t_{q-1}) A(0)\rangle
    =
    \sum_{\pi \in NC(q)} 
    \kappa_\pi (A(t_1) ... A(t_{q-1}) A(0)) \, ,
\end{equation}
here restricted to infinite temperature such that $\langle \, \cdot \, \rangle=\mathrm{tr}(\, \cdot \, )/D$ denotes the expectation w.r.t. the maximally mixed infinite-temperature state.
The summation runs over non-crossing partitions $\pi$ of the set $\{1, 2, ..., q\} \equiv [q],\: q \in \mathbb{N}$, denoted by $NC(q)$. 
A partition consists of disjoint subsets (blocks) $\{V\}$ and can be written as $\pi = \{ V_1, V_2, ..., V_{\# \pi}\}$, where $\# \pi$ is the number of blocks in $\pi$.
A partition is then said to be \emph{crossing} if we simultaneously have that (i) the elements $q_A, p_A$ and $q_B, p_B$ belong to distinct blocks $V_A$ and $V_B$, respectively, and that (ii) they are ordered as $p_A < q_A < p_B < q_B$. 

The thermal free cumulants factorize according to the blocks of the non-crossing partition $\pi$ as 
\begin{equation}\label{eq:cumulant_partition}
    \kappa_\pi(A_1 ... A_q)
    \equiv
    \prod_{V \in \pi}
    \kappa_{V}(A_1, ..., A_q),
\end{equation}
with for each block $V = \{i_1, ..., i_{|V|} \}$ of size $|V|$,
\begin{equation}\label{eq:cumulant_block}
    \kappa_{V}(A_1, ..., A_q)
    \equiv
    \kappa_{|V|}(A_{i_1} ... A_{i_{|V|}}).
\end{equation}

The definition~\eqref{eq:thermal_cumulants} is valid regardless whether the model is ergodic or not. As a concrete example, one can decompose the four-point correlation function with $q=4$ as
\begin{align}\label{eq:otoc_decomposition_1}
    \expec{A(t_3) A(t_2) A(t_1) A}
     = &\,
    \kappa_2 (t_2, t_1)\kappa_2 (t_3, 0)
    +
    \kappa_2  (t_2, t_3)\kappa_2 (t_1, 0)
    + \kappa_4 (t_1, t_2, t_3, 0),
\end{align}
which in diagrammatic form reads
\begin{equation}
    \expec{A(t_3) A(t_2) A(t_1) A}_\beta
    =
    \includegraphics[valign=c]{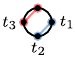}+
    \includegraphics[valign=c]{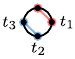}+
    \includegraphics[valign=c]{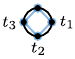}.
\end{equation}
Here we have fixed $\expec{A}=0$, which can be done without loss of generality, such that all terms containing $\kappa_1$ vanish.
In the diagrammatic representation we associate the operators at time $t_i$ to the vertices and connect vertices that belong to the same block of a partition by a straight line.
When full ETH is satisfied, the thermal free cumulants correspond to \emph{ETH cumulants}~\cite{pappalardiEigenstateThermalizationHypothesis2022}, which are defined as
\begin{equation}\label{eq:eth_cumulants}
      \kappa_q^{ETH}(\vec{t})
      \equiv
      \frac{1}{D}
      \sum_{i_1 \neq i_2 \neq .... \neq i_q}
      A(t_1)_{i_1 i_2}
      A(t_2)_{i_2 i_3}
      ...
      A(0)_{i_q i_1}\,,
\end{equation}
where the summation runs over different indices only. This particular form follows from the ETH-properties regarding the cacti and non-cacti.
Hereafter we will consider correlations along $\vec{t} = (t, 0, t, 0)$ and write all arguments in terms of $t$ only, which corresponds to the OTOC $\expec{A(t) A A(t) A}$.
This way, the OTOC decomposition in Eq.~\eqref{eq:otoc_decomposition_1} assumes the particularly simple form
\begin{equation}\label{eq:otoc_decomposition_free}
    \expec{A(t) A A(t) A}
    =
    2 \kappa_2 (t)^2
    +
    \kappa_4 (t),
\end{equation}
in terms of the free thermal cumulants.
We can relate the expected decomposition to ETH cumulants by expanding this OTOC in the eigenbasis, as 
\begin{equation}
    \expec{A(t) A A(t) A}
    =
    \frac{1}{D}
    \sum_{i j k l} e^{i (\omega_{ij} + \omega_{kl}) t}
    A_{ij} A_{jk}^* A_{kl} A_{li}^*\,
    .
\end{equation}
We can split this equation into different terms depending whether certain indices are equal to others 
\begin{equation}\label{eq:ETH_decomposition}
    \expec{A(t) A A(t) A}
    =
    \sum_{{\color{ForestGreen}i \neq j \neq k \neq l}} \dots + 
    \sum_{\color{RoyalPurple} \substack{{  i \neq j \neq l} \\ i = k}} \dots +
    \sum_{\color{RoyalPurple}  \substack{{i \neq j \neq k} \\ j = l}} \dots +
    \sum_{\color{BrickRed} \substack{{ i \neq j} \\ i = k,\: j = l}} \dots
\end{equation}
To simplify the discussion and as we justify in Appendix~\ref{app:diagonal_of_operator} for our concrete setting, we consider operators whose diagonal matrix elements $A_{ii}$ are zero.
This assumption allows for eliminating terms such as
\begin{equation}
    \sum_{i \neq j \neq k} A_{ii} A_{ij} A_{jk} A_{ki}
    =
    \includegraphics[valign=c]{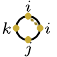}
    =
    0.
\end{equation}
In terms of the ETH diagrams, we thus have
\begin{align}\label{eq:OTOC_decomposition_graphical}
    \expec{A(t) A A(t) A}
    &=
    \includegraphics[valign=c]{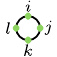}+
    \includegraphics[valign=c]{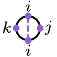}+
    \includegraphics[valign=c]{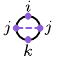}+
    \includegraphics[valign=c]{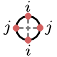} \\
    &=
    \kappa_4^{\mathrm{ETH}}(t) +
    2 \: \mathrm{cac}(t) +
    \mathrm{crossing}(t),\label{eq:OTOC_eth_diagram_decomposition}
\end{align}
with the fourth ETH cumulant, computed directly from definition~\eqref{eq:eth_cumulants}, given by
\begin{equation}
    \kappa_4^{\mathrm{ETH}}(t)
    =
    \includegraphics[valign=c]{images/otoc_diagram_1.pdf}
    =
    \frac{1}{D}
    \sum_{i \neq j \neq k \neq l}
    e^{i (\omega_{ij} + \omega_{kl}) t}
    A_{ij} A_{jk}^* A_{kl} A_{li}^*,
\end{equation}
 and
\begin{equation}\label{eq:cactus}
    \mathrm{cac}(t)
    =
    \includegraphics[valign=c]{images/otoc_diagram_2.pdf}
    =
    \includegraphics[valign=c]{images/otoc_diagram_3.pdf}
    =
    \frac{1}{D}
    \sum_{i \neq j \neq k}
    e^{i (\omega_{ij} + \omega_{ik}) t}
    |A_{ij}|^2 |A_{ik}|^2
\end{equation}
denotes the two cacti diagrams appearing in Eq.~\eqref{eq:OTOC_decomposition_graphical}.
Since we take $t_3 = t_1 = t$ and $t_2 = 0$ both contributions are identical.
Whenever we mention to the cactus diagram in our model, we will be referring to Eq.~\eqref{eq:cactus}.
Similarly, we have the single crossing contribution for the OTOC, which will also be a central quantity in our discussion:
\begin{equation}\label{eq:crossing}
    \mathrm{crossing}(t) 
    = 
    \includegraphics[valign=c]{images/otoc_diagram_4.pdf}
    =
    \frac{1}{D}
    \sum_{i \neq j} e^{i \omega_{ij} t}|A_{ij}|^4.
\end{equation}

Note, however, that Eq.~\eqref{eq:OTOC_eth_diagram_decomposition} is a completely general decomposition, and no stronger assumptions, except for $A_{ii}=0$ are made here. 
One can always compute the corresponding diagrams in any scenario, both for ergodic and integrable models.
The question is whether in the thermodynamic limit, (i) the cactus diagram~\eqref{eq:cactus} factorizes into a product of second order cumulants as $\mathrm{cac}(t) \rightarrow [\kappa_2^{\mathrm{ETH}}(t)]^2$ and (ii) if the crossing partition~\eqref{eq:crossing} vanishes, as predicted by full ETH.


\section{The dual-unitary XXZ circuit}\label{sec:DU_circuits}
We will consider these probes of full ETH in integrable brickwork quantum circuits that arise as the Trotterization of the integrable XXZ Hamiltonian.
Specifically, we consider two-qubit unitary matrices of the form
\begin{equation}\label{eq:XXZ_parametrization}
    U 
    = 
    \exp\left[-i \tau
    \left( X \otimes X 
    +  Y \otimes Y 
    + J_z\, Z \otimes Z 
    \right)\right],
\end{equation}
with $X,Y,Z$ the Pauli matrices, $J_z$ the anisotropy, and $\tau \in \mathbb{R}$ the Trotter step. We represent this unitary graphically as
\begin{equation}
    U = \includegraphics[valign=c]{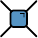}\,.
\end{equation}
For a one-dimensional lattice of $L$ qubits, we fix the architecture of our unitary circuit with open boundary conditions, with the product of even and odd layers, as follows: 
\begin{equation}\label{eq:floquet_circuit}
    U_F = U_e U_o = (\id \otimes U \otimes ... \otimes U)(U \otimes ... \otimes U \otimes \id).
\end{equation}
This circuit can be represented graphically as
\begin{equation}\label{eq:floquet_circuit_drawn}
    U_F = \includegraphics[valign=c]{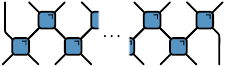}.
\end{equation}
For $t$ discrete time steps, the unitary evolution operator is given by $(U_F)^t$.
The resulting dynamics is integrable for any choice of the Trotter step $\tau$ and anisotropy $J_z$, supporting an extensive set of conserved charges and admitting an exact construction of the eigenstates of $U_F$ through the Bethe ansatz~\cite{vanicat_integrable_2018,ljubotina_ballistic_2019,friedman_spectral_2019,krajnik_integrable_2020,claeys_correlations_2022}.

We first focus on the dual-unitary point $\tau = \pi/4$, where analytical progress can be made. 
Since the exact solution for the eigenstates has not previously appeared in the literature and is of independent interest, we give it here in full detail. Readers interested in the results for full ETH can skip to Sec.~\ref{subsec:results_DUXXZ}.
In order to simplify the discussion and our notation, we rewrite the parametrization introduced in Eq.~\eqref{eq:XXZ_parametrization} in a more direct form.
At the dual-unitary point we can re-express the gates as $U = \CPhase \swap$, where
\begin{equation}
    \swap  = \swapdraw    
\end{equation}
denotes the swap gate and its diagrammatic representation and 
\begin{equation}
    \CPhase 
    = 
    \ketbra{0}{0} \otimes \id +  \ketbra{1}{1} \otimes \left(\ketbra{0}{0} + e^{i \phi}\ketbra{1}{1}\right) 
    =
    \cphasedraw
\end{equation}
is a controlled-phase gate expressed in the computational basis. This representation highlights the dual-unitarity of the gate, where $U$ remains unitary under a reshuffling of indices corresponding to swapping time evolution by space evolution, since the gate has an identical representation in the temporal~ (vertical) and spatial (horizontal) direction~\cite{gopalakrishnan_unitary_2019,bertiniExactCorrelationFunctions2019}. 
To avoid pathological cases we take the phase $\phi$ incommensurate with $\pi$.
This circuit yields an integrable Trotterization of the XXZ Hamiltonian with Trotter step at the dual-unitary point $\tau = \pi/4$, 
and the phase $\phi$ can be rewritten in terms of the anisotropy $J_z$ upon reparametrization~\cite{suzukiComputationalPowerOne2022b}.
In this limit the Floquet circuit can be written as
\begin{equation}\label{eq:circuit_int}
    U_F
    =
    \includegraphics[valign=c]{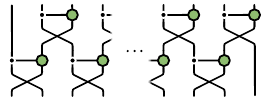}.
\end{equation}
The non-interacting limit of the circuit $U$ is obtained by setting $\phi=0$. In this limit the controlled phase gate is the identity and the full circuit is built from swap gates only and is denoted by $\mathcal{S}$. 
Consequently, $\mathcal{S}$ implements a generalized shift of the lattice sites, for which it can be directly checked that $\mathcal{S}^L = \id$.

\subsection{Solitons and charges in DU circuits}
\label{subsec:solitons}

\begin{figure}
    \centering
    \includegraphics[width=0.66\columnwidth]{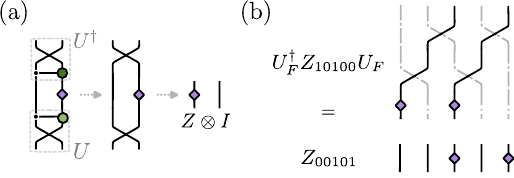}
    \caption{
    (a) Action of the DU XXZ gate on solitons. The darker green shade represents the conjugate phase in the gate $U^\dagger$.
    (b) Illustration of the action of the SWAP circuit on the soliton $Z_{10100}$ for $L = 5$.}
    \label{fig:soliton_panel}
\end{figure}

The particular combination of dual-unitarity and integrability strongly constrains the dynamics, since it implies the existence of solitons: traceless operators which, when conjugated by the Floquet DU circuit, are simply shifted along the lattice (up to a phase)~\cite{bertini_operator_2020}. 
These solitons are both important in classifying the exact eigenstates and strongly constrain the matrix elements of operators composed of solitons.
For the dual-unitary XXZ circuit of Eq.~\eqref{eq:circuit_int}, the solitons correspond to the Pauli $Z$-matrices, since
\begin{equation}
    U_F^\dagger Z_x U_F
    =
    Z_{x + y},
\end{equation}
with $x, y \in \mathbb{Z}$, as illustrated in Fig.~\ref{fig:soliton_panel}. Graphically,
\begin{equation}
    Z_x
    =
    \hdots
    \underset{x - 1}{\leg}
    \:
    \underset{x}{\soliton}
    \:
    \underset{x + 1}{\leg}
    \hdots,
\end{equation}
with $Z$ being the Pauli matrix acting on the computational basis states as $Z\ket{0}=-\ket{0}$ and $Z\ket{1}=\ket{1}$. 
In our particular case, outside the boundaries we have $y = \pm 2$,  corresponding to right- and left- ballistic propagation respectively.
This implies that the circuit in Eq.~\eqref{eq:floquet_circuit} is, from the point of view of solitons, equivalent to the phase-free SWAP circuit $\mathcal{S}$. 
Correlation functions between solitons, including two-point functions $\expec{A(t)A}$ and OTOCs, therefore become periodic in time~\cite{rampp_hayden-preskill_2024}, preventing the dynamics from being mixing. 
In particular, in contrast to chaotic models, time-evolved observables do not become freely independent from static observables even at late times.
Instead, solitons $Z_x(t)$ are supported at different lattice sites than  $Z_x(0)$ unless $t$ is an integer multiple of $L$ and hence arbitrary infinite temperature correlation functions factorize. For instance the OTOC factorizes as $\langle Z_x(t)Z_xZ_x(t)Z_x\rangle = \langle Z_x^2 \rangle^2$ and is constant for most times for finite systems and for all times in the thermodynamic limit. This factorization of joint correlators between $Z_x(t)$ and $Z_x(0)$ implies their classical independence, incompatible with free independence.

Conserved charges can be systematically constructed from solitons, and there is one-to-one correspondence between solitons and charges in DU circuits~\cite{holden-dyeFundamentalChargesDualunitary2023}. 
A simple example is the total magnetization, which can be constructed from all single-body solitons:
\begin{equation}\label{eq:magnetization_op}
\begin{split}
    S_z
    =
    \sum_x Z_x
    =
    \soliton\leg\hdots\leg + 
    ~ ... ~ +
    \leg \leg\hdots\soliton.
\end{split}
\end{equation} 
In order to construct higher-body conserved charges, we index these charges by a bitstring $a = a_1 a_2 ... a_L$ of size $L$, with $a_i = 0, 1$. The number of non-zero bits $|a| = \sum a_i \in \mathbb{N}$ will describe either the number of up-spins in a {state} or, when associated with an {operator}, the number of solitons. 
An arbitrary $|a|$-body soliton, given a bitstring $a$, is written as:
\begin{equation}\label{eq:soliton_bitstring}
    Z_{(a)}
    \equiv
    \bigotimes_{x = 1}^{L}
    Z^{a_x}_x.
\end{equation}
That is, we have the identity at site $x$ whenever $a_x = 0$ and $Z_x$ otherwise. 
By doing so, we can write down conserved charges from a bitstring $a$ for the circuit as:
\begin{equation}\label{eq:charge_bistring}
    Q_{(a)} 
    = \sum_{t = 0}^{L - 1} \mathcal{S}^{-t} Z_{(a)} \mathcal{S}^t
    = \sum_{t = 0}^{L - 1}  Z_{\sigma^t(a)}.
\end{equation}
Here we have introduced the corresponding permutation function $\sigma \in S_L$. 
For $n = 1, 2,..., L$ this permutation acts on a bitstring $a$ as $\sigma(a_2) = a_1$ and $\sigma(a_L) = a_{L-1}$ at the edges and in the bulk
\begin{equation}\label{eq:permutation_def}
    \sigma(a_x) =
    \begin{cases}
        a_{x + 2}, & \text{for $x$ odd},\\
        a_{x - 2}, & \text{for $x$ even}.
    \end{cases}
\end{equation}
We have explicitly assumed $L$ to be odd, as also required for the particular architecture illustrated in Eq.~\eqref{eq:floquet_circuit_drawn}.
Furthermore, 
note that we have $\sigma^L(a) = a$, consistent with $\mathcal{S}^L = \id$.
It directly follows that $[\mathcal{S}, Q_{(a)}] = 0$ for any bitstring $a$ and that these charges are indeed conserved quantities. As an example, for $a = 10100$ the corresponding charge reads
\begin{equation}\label{eq:soliton_visual}
\begin{split}
    Q_{10100}
    =& \soliton ~ \leg ~ \soliton ~ \leg ~ \leg +
    \leg ~ \leg ~ \soliton ~ \leg ~ \soliton +
    \leg ~ \leg ~ \leg ~ \soliton ~ \soliton \\
    & 
    + \leg ~ \soliton ~ \leg ~ \soliton ~ \leg 
    + \soliton ~ \soliton ~ \leg ~ \leg ~ \leg.
\end{split}
\end{equation}  
Because of the resulting superextensive number of conserved charges, these models are also referred to as superintegrable~\cite{gombor_superintegrable_2022,gombor_integrable_2024}. By the above construction, charges constructed from bitstrings $a_1$ and $a_2$ in the same orbit of $\sigma$, i.e., $\sigma^l(a_1)=a_2$ for some $l$ give rise to the same charge.

States in the computational basis are eigenstates of the solitons by construction, and the corresponding eigenvalues can be directly obtained. 
Consider the bitstring $m = m_1 m_2 ... m_L$ which we associate with the state $\ket{m} = \ket{m_1 ... m_L}$, then
\begin{equation}
    Z_{(a)}
    \ket{m_1 ... m_L}
    =
    (-1)^{|a| + a \cdot m}
    | m_1 m_2 ... m_L  \rangle,
\end{equation}
The quantum number $\lambda^m_a$ corresponding to the charge $Q_{(a)}$ and eigenstate $\ket{m}$, such that $Q_{(a)} | m \rangle = \lambda^m_a | m \rangle$, hence reads
\begin{equation}\label{eq:charge_eigenvalue}
    \lambda^m_a 
    =
    (-1)^{|a|}
    \sum_{t = 0}^{L - 1}
    (-1)^{\sigma^t({a}) \cdot {m}},    
\end{equation}
where $\sigma^t(a)$ is a $t$-fold permutation of $a$. 
States for which all quantum numbers are the same belong to the same (irreducible) sector.

There is an equivalence between the role of the state and the generating soliton when computing a quantum number through Eq.~\eqref{eq:charge_eigenvalue}.
This equivalence follows from the simple observation that 
$
    \sigma^t({a}) \cdot {m}
    =
    {a} \cdot \sigma^{-t}({m}),
$
such that we can rewrite Eq.~\eqref{eq:charge_eigenvalue} as
\begin{align}\label{eq:soliton_state_duality}
    \lambda^m_a
    &=
    (-1)^{|a|}
    \sum_{t = 0}^{L - 1}
    (-1)^{\sigma^t({a}) \cdot {m}} 
    =
    (-1)^{|a|}
    \sum_{t = 0}^{L - 1}
    (-1)^{{a} \cdot \sigma^{-t}({m})} \nonumber\\
    &=
    (-1)^{|a|}
    \sum_{t = 0}^{L - 1}
    (-1)^{{a} \cdot \sigma^{t}({m})} = (-1)^{|a|-|m|} \lambda^a_m,
\end{align}
since $\sigma^{-1} = \sigma^{L - 1}$, $\sigma^{-2} = \sigma^{L - 2}$ and so on. 
The RHS hence returns $\lambda^{a}_m$, i.e. the inverse scenario in Eq.~\eqref{eq:charge_eigenvalue}, where $m$ plays the role of the soliton and $a$ plays the role of the state. 
We thus arrive at
$
    (-1)^{|m|} 
    \lambda^{(a)}_m
    =
    (-1)^{|a|} 
    \lambda^{(m)}_a.
$
By taking $|a| = |m|$ we write down the identity which we call the soliton-state duality (similar to the string-charge duality in more general interacting integrable systems~\cite{ilievski_string-charge_2016}), namely $\lambda^{a}_m = \lambda^{m}_a$. That is, within the same magnetization sector, the quantum number $\lambda^{a}_m$ is invariant upon swapping $a$ and $m$.

\subsection{Eigenvalues and eigenstates}
\label{subsec:eigenstates}
The eigenstates of the unitary evolution operator \eqref{eq:circuit_int} can be obtained in a similar manner. In order to obtain these eigenstates we note that, starting from a state $\ket{m}$, the unitary evolution only generates states in the orbit $\{|m\rangle, |\sigma(m)\rangle, \dots,|\sigma^{L-1}(m)\rangle\}$ up to phases due to the interaction. A similar observation was used in Ref.~\cite{gopalakrishnan_facilitated_2018} to construct nonthermal eigenstates for quantum cellular automata.
It is a direct check using Eq.~\eqref{eq:charge_eigenvalue} that all states within this orbit have identical quantum numbers, as should be the case. 
The reverse also holds: using the soliton-state duality [Eq.~\eqref{eq:soliton_state_duality}] it can be shown that if two states $\ket{m}$ and $\ket{n}$ have the same quantum numbers, they necessarily lie in the same orbit. 
Assuming $L$ to be prime from here on, all orbits with the exception of the fully polarized states have length $L$.
In Appendix~\ref{app:charge_sector} we discuss this assumption and how the following calculations have to be adapted for composite $L$.
The full Hilbert space consequently decomposes in two blocks of size $1\times 1$ and $(2^L-2)/L$ blocks of size $L \times L$. We hence need a total of $(2^L - 2)/L \sim 2^L/L$ quantum numbers to fully characterize the orbit or, equivalently, the sector in the Hilbert space, consistent with the number of conserved charges.

\begin{figure}
    \centering
    \includegraphics[width=0.35\columnwidth]{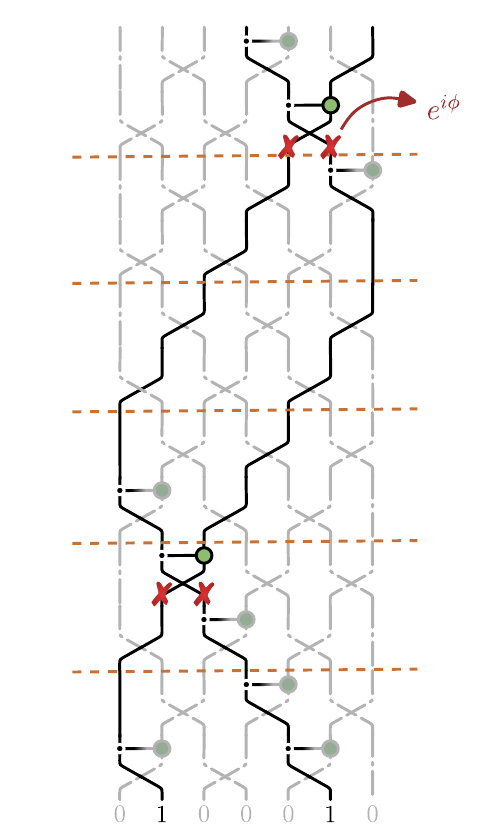}
    \caption{Illustration of how two spins interact twice along the orbit of $\ket{\sigma^t(m)}$, as highlighted by the red marks. 
    The two spins will scatter each other whenever one of the is right moving and the other one is left moving. Here they meet at $t = 2$ and $t = 6$. 
    For visual clarity, we omit the controlled-phase gates whenever the control qubit is zero.}
    \label{fig:scatter_illustration}
\end{figure}

The unitary $U_m$ obtained by projecting the full circuit $U_F$ onto the sector defined by an initial state $m$ acts as $U_m \ket{\sigma^s(m)} = e^{i \phi_s }  \ket{\sigma^{s + 1}(m)}$.
The projected unitary for any of the sectors hence follows as
\begin{equation}\label{eq:sector_unitary_comp}
    U^{(m)}
    =
    \begin{matrix}
    \begin{pmatrix}
    0 & 0 & 0 & 0 & \hdots & 0 & e^{i \phi_L} \\ 
    e^{i \phi_1} & 0 & 0 & 0 & \hdots & 0 & 0 \\ 
    0 & e^{i \phi_2} & 0 & 0 & \hdots & 0 & 0 \\ 
    \vdots & \vdots & \vdots & \vdots & \ddots & \vdots & \vdots \\ 
    0 & 0 & 0 & 0 & \hdots & e^{i \phi_{L-1}} & 0
    \end{pmatrix}
    \end{matrix}\,.
\end{equation}  
The individual phases $\phi_s$ can be systematically calculated within every orbit but they depend on the choice of initial state, as discussed in Appendix~\ref{app:scattering_matrix}.
We can nevertheless collect these phases into a full `geometric' phase by performing a basis transformation as
\begin{equation}\label{eq:scattering_phase}
  |\tilde{s}\rangle = e^{i \Phi_s} |\sigma^s(m)\rangle \qquad \textrm{with} \quad \Phi_s = \sum_{j = 0}^s \phi_j,
\end{equation}
where $\Phi_0 = 0$. We will refer to $\Phi_s$ as the \emph{partial scattering phase} and $\Phi_L$ as the \emph{total scattering phase}. 
Within this basis, the unitary evolution operator acts as
\begin{equation}\label{eq:sector_unitary_transf}
    \tilde{U}^{(m)}
    =
    \begin{matrix}
    \begin{pmatrix}
    0 & 0 & 0 & 0 & \hdots & 0 & e^{i \scatteringphase_L} \\ 
    1 & 0 & 0 & 0 & \hdots & 0 & 0 \\ 
    0 & 1 & 0 & 0 & \hdots & 0 & 0 \\ 
    \vdots & \vdots & \vdots & \vdots & \ddots & \vdots & \vdots \\ 
    0 & 0 & 0 & 0 & \hdots & 1 & 0
    \end{pmatrix}
    \end{matrix}\,.
\end{equation}
depending only on the total scattering phase.
This total scattering phase can be directly obtained by noting that two distinct spins in the state $\ket{m}$ ``scatter'' at each other exactly twice along the orbit $\{\ket{\sigma^s(m)}\}_{s=0,...,L-1}$, as illustrated in Fig.~\ref{fig:scatter_illustration}, such that the total scattering phase is uniquely fixed by the number of spin excitations as
\begin{align}
    \scatteringphase_L = 2|m|(|m|-1)\phi \, .
\end{align}
The resulting matrix \eqref{eq:sector_unitary_transf} represents a unidirectional hopping model with twisted boundary conditions. The corresponding eigenstates follow as plane waves of the form
\begin{equation}\label{eq:eigenstates}
    \ket{\varepsilon_k^m}
    =
    \frac{1}{\sqrt{L}}
    \sum_{s = 0}^{L - 1}
    \exp[ i (s \theta_k^{(m)} + \Phi_s)]
    \ket{\sigma^s(m)},
\end{equation}
where the eigenphases satisfy the quantization condition
\begin{equation}\label{eq:eigenphases}
    \theta_k^{(m)}
    =
    \frac{2\pi k - \scatteringphase_L}{L}, \qquad k = 0,1 \dots L-1.
\end{equation}
This equation corresponds to the Bethe equation for this model.
The eigenvalues of the mode follow as the $L$ roots of unity, up to a overall phase shift given by the angle $\scatteringphase_L$. 
The resulting spectrum of the evolution operator is highly degenerate as the eigenvalues depend only on the magnetization via the total scattering phase and the quantum number $k$.
More precisely there are $\binom{L}{|m|}$ states with same magnetization $|m|$. In the case $0 < |m| < L$,  there are also $L$ states which are generalized shifts of a single representative state $m$.
This means that the total scattering phases are $\binom{L}{|m|}/L$-fold degenerate.
As each eigenphase additionally only depends on the momentum quantum number $k$, eigenphases corresponding to the same total scattering phase and the same momentum quantum number are $\binom{L}{|m|}/L$-fold degenerate as well.

Note that while the eigenvalues depend only on the total scattering phase, the eigenvectors obtained above do depend on the partial scattering phase, making them significantly more complicated.
In Appendix~\ref{app:fkm_algorithm} we discuss how to efficiently generate the representative bitstrings for solitons and states, relevant for Eqs.~\eqref{eq:charge_bistring} and~\eqref{eq:eigenstates} when constructing the charges and eigenstates. 

Despite being presented here for the Trotterized XXZ Hamiltonian, the above diagonalization procedure can be extended to arbitrary qubit circuits hosting left and right moving solitons, where these solitons are additionally required to be the same in both directions in case of open boundary conditions.

\begin{figure}[t]
    \centering
    \includegraphics[width=0.66\columnwidth]{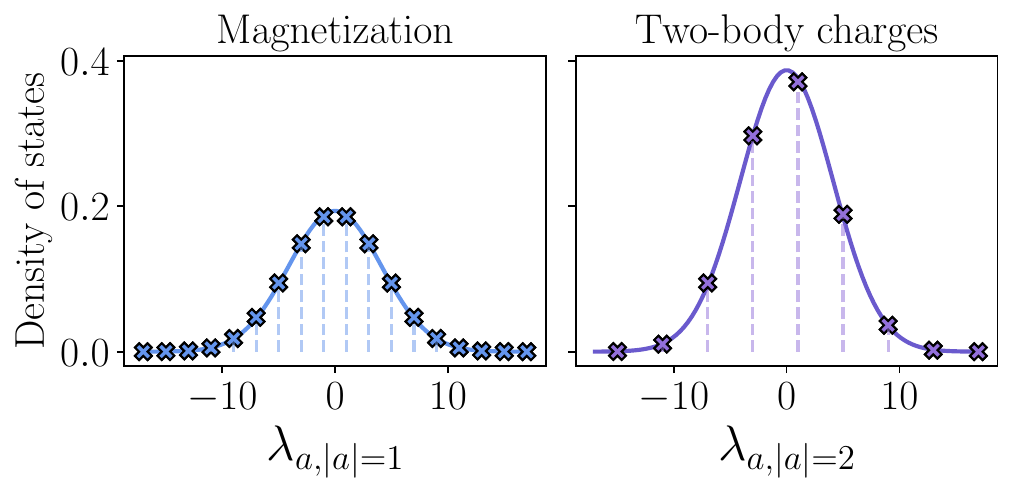}
    \caption{
    Distribution of eigenvalues for the one-body charge (magnetization) and two-body charges for $L = 17$. 
    The former is binomially distributed as ${L \choose l}$ with $l = 0, 1, ..., L - 1, L$, while the latter follows a modified binomial distribution $2{L \choose l}$ with $l = 1, 3, ..., L - 2, L$.
    }
    \label{fig:charge_distribution}
\end{figure}

\subsection{Crossing diagrams and factorization}
\label{subsec:results_DUXXZ}

Returning to probes of full ETH, by using these exact eigenstates we can explicitly evaluate the cactus diagrams~\eqref{eq:cactus} and the crossing contribution~\eqref{eq:crossing} for observables that corresponds to a soliton.
Within a single sector we explicitly relate the crossing contribution to the charge quantum numbers defining that sector, while for the full many-body Hilbert space these can be expressed in terms of statistical moments of the charge distribution over the computational basis states.

Let us first consider the crossing contribution in a fixed sector labeled by the representative state $m$, for a local observable $Z_x$ given by a soliton at site $x$:
\begin{align}
    C_m = \frac{1}{L}\sum_{k \neq l}\left|\bra{\varepsilon_k^m}Z_x\ket{\varepsilon_l^m}\right|^4.
\end{align}
We redefine $Z_x$ to have vanishing diagonal elements in the computational basis, which here corresponds to a shift by the total magnetization (see Appendix~\ref{app:diagonal_of_operator}) and simplifies the following analysis without modifying the results for the crossing contribution.
The choice of soliton as an observable also guarantees that (i) the observable only has nonzero matrix elements between states in the same sector, i.e. there are selection rules, and (ii) the matrix elements will not depend on the partial scattering phases since this operator is diagonal in the computational basis and all nontrivial phases will hence cancel. 
As shown in Appendix~\ref{app:crossing_DU}, the resulting crossing contribution can be expressed purely in terms of the associated eigenvalues of the one-body and two-body conserved charges as 
\begin{equation}\label{eq:crossing_soliton_basis}
C_m = \frac{2}{L^3}\sum_{a,\, |a|=2} \left(\lambda_a^m\right)^2 - \frac{M^4}{L^4} + \frac{1}{L}\, .
\end{equation}
Here,  $M = 2|m| - L = \lambda_{a,\, |a|=1}$ denotes the magnetization, i.e. the unique charge constructed from the single-body solitons.
The contribution within each sector can be either vanishing or not, depending on both the two-body conserved charges and the one-body conserved charge. For ergodic systems we expect the crossing contribution to be suppressed in the total magnetization sector, which would here correspond to the crossing contribution being suppressed in most (soliton) sectors.

\begin{figure}
    \centering
    \includegraphics[width=0.66\columnwidth]{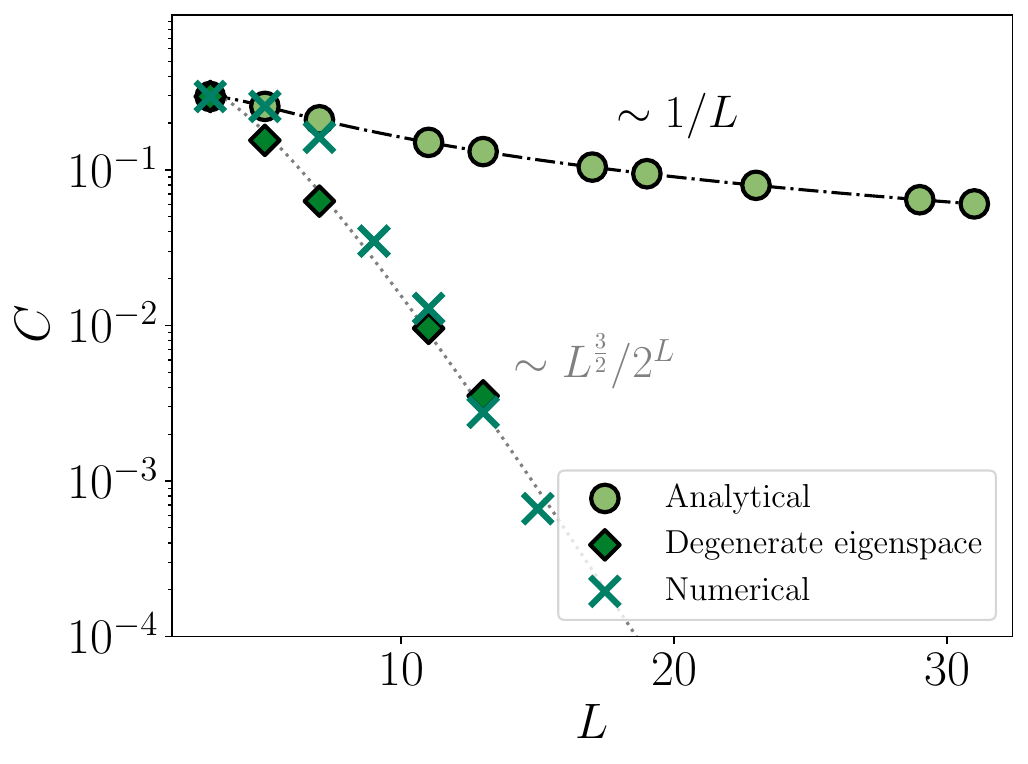}
    \caption{
        Crossing contribution for the observable $Z$ at the middle of the circuit.
        The dot dashed line denotes the exact analytical prediction from Eq.~\eqref{eq:crossing_analytical}, which decays polynomially with $L$.
        The dotted gray line shows the quasi-exponential scaling.
        Circular markers: crossing diagram for the unitary in block-diagonal form, computed from the average of each of the $L \times L$ sectors,  based on the analytical eigenstates~\eqref{eq:eigenstates}.
        Diamond markers: Crossing contributions calculated in the basis constructed from the degenerate eigenspace.
        Crosses: Crossing contributions calculated in the basis obtained by numerical diagonalization over the full Hilbert space.
    }
    \label{fig:crossing_analytical}
\end{figure}

The crossing contribution $C$ for the full Hilbert space follows as
\begin{align}\label{eq:crossing_average}
    C 
    & = 
    \frac{1}{D} \sum_{\textrm{ Sectors}\, m,n} \sum_{(m,k) \neq (n,l)} \left|\bra{\varepsilon_k^m}Z_x\ket{\varepsilon_l^n}\right|^4 \nonumber \\
    & = 
    \frac{1}{D} \sum_{\textrm{ Sectors}\, m} \sum_{k \neq l} \left|\bra{\varepsilon_k^m}Z_x\ket{\varepsilon_l^m}\right|^4 \nonumber \\
    & = 
    \frac{L}{D}\sum_{\text{Sectors}\, m} C_m
     = 
    \big\langle C_m \big\rangle_{\text{Sectors}\, m} \, .
\end{align}
The full crossing contribution is thus given by the average of the crossing contribution over all sectors. 
Following Eq.~\eqref{eq:crossing_soliton_basis} this amounts to computing the fourth moment of the magnetization $M$ and the second moment for the distribution of two-body charges $\lambda_{a,\, |a|=2}$.
These different contributions can both be evaluated explicitly.
\begin{align}
\left\langle M^4\right\rangle_{\text{Sectors}\, m}= 3L^2 - 2L\,,\quad 
\left\langle \lambda_{a,\, |a| = 2}^2\right\rangle_{\text{Sectors}\, m}= L\,.
\end{align}
These expectation values can be calculated by using that both are binomially distributed, as illustrated in Fig.~\ref{fig:charge_distribution} and explicitly shown in Appendix~\ref{app:charge_distribution}. Using that there are approximately $2^L/L$ sectors (neglecting the two sectors of size one corresponding to the fully polarized states, which do not contribute to the crossing diagrams), and that there are $(L - 1)/2$ unique two-body charges, we get an exact expression which depends only on the system size:
\begin{equation}\label{eq:crossing_analytical}
    C = \frac{2}{L} - \frac{4}{L^2} + \frac{2}{L^3}.
\end{equation}
We see that the crossing here is suppressed polynomially as $O(1/L)$, in stark contrast to the chaotic regime where we expect an exponential scaling.
Moreover, this $O(1/L)$ suppression upon averaging over all sectors indicates that most of the individual sectors show the same $O(1/L)$ scaling, recovering the prediction of full ETH within most sectors. 
This result is illustrated in Fig.~\ref{fig:crossing_analytical}, where a calculation of the crossing contributions based on the eigenstates~\eqref{eq:eigenstates} shows the expected $1/L$ crossing.

However, in practice we observe a distinct behavior in more generic scenarios.
In all results so far, we focused on the analytical eigenstates given by Eq.~\eqref{eq:eigenstates}, our so-called ``soliton basis''.
These eigenstates directly led to the analytical expression in Eq.~\eqref{eq:crossing_analytical} above.
However, \emph{in the presence of degeneracies the crossing contribution [Eq.~\eqref{eq:crossing}] is basis-dependent}
\footnote{
This is of course the case for ETH diagrams in general, not only the crossing contributions.
}, and the degeneracies in the dual-unitary XXZ model allow for the construction of eigenstates that are linear superpositions of the states from Eq.~\eqref{eq:eigenstates}.
The question which remains is: what could one then expect in a more generic basis constructed from this degenerate subspace?
In this case, we see that this new set of eigenstates leads to a very different result compared to our initial analytical prediction~\eqref{eq:crossing_analytical}, showing a stronger suppression of the crossing diagrams (as illustrated in Fig.~\ref{fig:crossing_analytical}).
Note that these states are however no longer exact eigenstates of the conserved charges \eqref{eq:charge_bistring}.
This new set of eigenstates living in the degenerate subspace, which is exponentially large as we will argue below, gives rise to this distinct and much stronger suppression.
This stronger suppression in fact naturally arises in numerical experiments where we directly perform exact diagonalization of the unitary circuit from Eq.~\eqref{eq:floquet_circuit} and compute the crossing contributions based on these numerical eigenstates. Numerical solvers do not ``know'' about the soliton basis~\eqref{eq:eigenstates}, but rather yield numerical eigenstates corresponding to an effectively random linear combination of degenerate soliton eigenstates.

Specifically, as discussed in the previous subsection, the quasi-energies in Eq.~\eqref{eq:eigenphases} are uniquely labeled by their magnetization and `momentum' $k$.
Namely, the eigenstates in a given $L \times L$ sector are uniquely distinguished by the $L$ different momenta they can assume.
Meanwhile, there are $\frac{1}{L}{L \choose |m|}$ degenerate eigenstates with same magnetization and momenta, which belong however to different sectors.
For large $L$, this establishes a degenerate eigenspace of dimension $\frac{1}{L}{L \choose |m|} \approx 2^L/L^{3/2}$, which we call \emph{quasi-exponential} and which is exponentially larger than the $L \, \times \, L$ sectors constructed from the soliton basis. 
Constructing eigenstates as random superpositions of the presented eigenstates, this choice of basis effectively implements random rotations with this dimensionality~\cite{foiniEigenstateThermalizationHypothesis2019a}. 
This choice of eigenstates results in a quasi-exponential scaling of the crossing contribution as $L^{\frac{3}{2}}/2^L$, since for random rotations the crossing contributions are suppressed as the inverse of the dimension of the rotation matrix -- an initial motivation for full ETH~\cite{maillard_high-temperature_2019}.
An explicit random matrix calculation considering all sectors provides a a similar polynomial correction of $L^2$, as we discuss in Appendix~\ref{app:random_basis}.

\begin{figure}
    \centering
    \includegraphics[width=0.66\columnwidth]{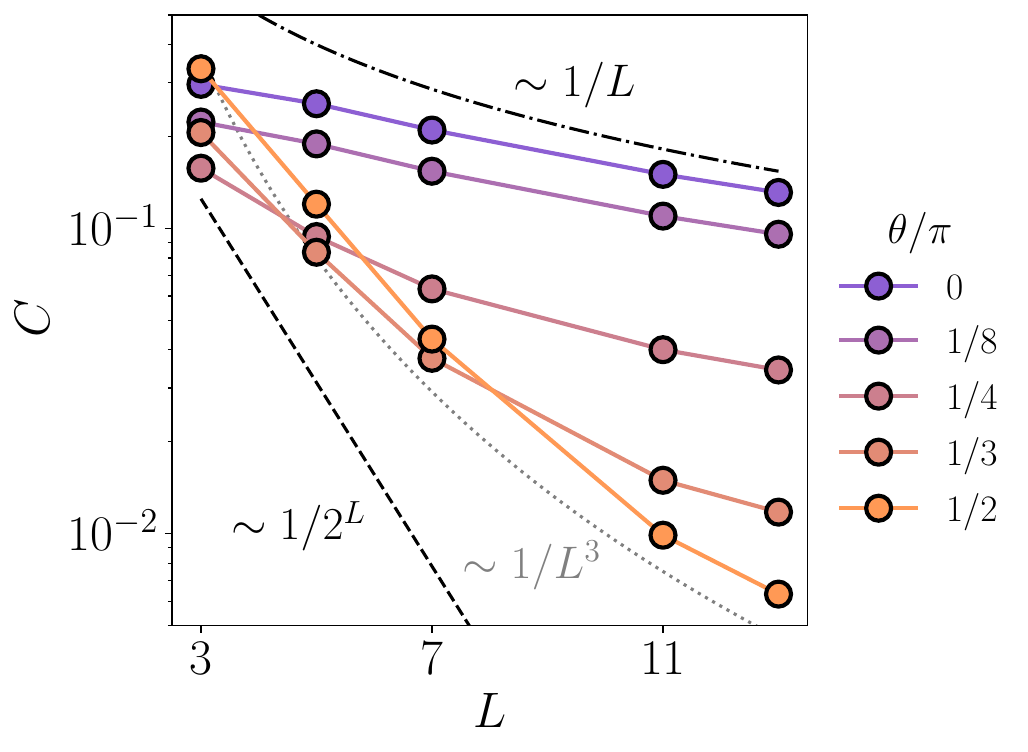}
    \caption{Crossing contribution and factorization of the cactus diagrams in the soliton basis as a function of $L$ for different observables $A = Z_x \cos \theta + X_x \sin \theta$.
    The numerics indicate a power-law scaling $1/L^\alpha$ in all cases, with increasing exponent $\alpha \geq 1$ for larger $\theta$.
    }
    \label{fig:crossing_fac_analytical_observable}
\end{figure}

This suppression is consistent with the numerical results of Fig.~\ref{fig:crossing_analytical}.
We plot the analytical expression from Eq.~\eqref{eq:crossing_analytical} as the black dot-dashed line and the quasi-exponential scaling of $L^{\frac{3}{2}}/2^L$ as dotted gray line.
Numerical results are indicated by markers, where our analytical results allow for the determination of the crossing contribution in the eigenbasis \eqref{eq:eigenstates} up to $L = 31$.
First, we use the analytical eigenstates to numerically compute the crossing partition for each $L \times L$ sector as Eq.~\eqref{eq:crossing_soliton_basis}. 
Afterwards, by following Eq.~\eqref{eq:crossing_average} we take the numerical average of all these sector-wise contributions.
In the second scenario, we construct a random superposition of degenerate eigenstates, using their analytical form in Eq.~\eqref{eq:eigenstates}.
We then compute, numerically, the crossing contribution in this new basis. 
These are shown with the diamond-shaped markers in Fig.~\ref{fig:crossing_analytical} and we can see that they follow the quasi-exponential scaling.
Similarly, with the crosses we plot the purely numerical results computed from the full Hilbert space, exactly as done in Fig.~\ref{fig:crossing_scaling}, confirming that we indeed observe a quasi-exponential scaling there as well.
This plot shows that the polynomial scaling of $1/L$ is a feature of this model at this very particular basis, constructed directly from the soliton formalism. 
We refer to this as the soliton basis.
However, as soon as degeneracies are present it is then possible to construct a new set of eigenstates which provide a very different scaling.
In this case, the scaling will depend the dimensionality of the largest magnetization sector, and we find the quasi-exponential scaling. 
Fully numerical results will give rise to the same scaling, since the eigenstates we obtain numerically will not, in general, agree with simple eigenstates obtained through the soliton-based diagonalization procedure.
Instead, numerical procedures will yield eigenstates as random superpositions of degenerate eigenstates in the soliton basis.

These results are particular to observables that consist of solitons.
We now consider a more general observable which is a superposition of $Z_x$ and $X_x$,
$A = Z_x \cos \theta + X_x \sin \theta$.
In this way we interpolate from $Z$ with $\theta = 0$, which couples only states with same magnetization, to $X_x$ with $\theta = \pi/2$, which couples states whose magnetization differ by one.
We consider the crossing contribution in the soliton basis, with results shown in Fig.~\ref{fig:crossing_fac_analytical_observable}.
First, we observe the $1/L$ scaling for $\theta = 0$.
By increasing $\theta$ we can see that the crossing contributions are also stronger suppressed, decaying more sharply in the case where $A = X_x$.
Nevertheless, we observe a power-law scaling for all choices of observables.
In general, these numerical results suggest that for $\theta > 0$, i.e. observables other than $Z$, the crossing contribution decays as $1/L^\alpha$ with $\alpha > 1$.
These results are consistent with the observation that, since $X_x$ couples different sectors, we are effectively working with a Hilbert space of larger dimensions, such that the matrix elements in this basis will undergo a stronger suppression compared with what we observe for $A = Z_x$, which does not couple different sectors.

In general it is also necessary to verify the validity of the factorization of the cacti in the form of Eq.~\eqref{eq:eth_factorization} in order to properly probe full ETH in a given model.
In our case, however, the cacti and the crossing diagrams are related and the vanishing of the crossing diagrams implies factorization.
This equivalence is due to the fact that the soliton operators square to the identity, for which observables it follows that $\mathrm{cac}(t = 0) = \kappa_2^{ETH}(t=0)^2 - C(t = 0)$, as shown in Ref.~\cite{pappalardiGeneralEigenstateThermalization2025}.
For the OTOC decomposition of Eq.~\eqref{eq:OTOC_decomposition_graphical} our analysis hence provides a complete picture and it is sufficient to only consider the crossing contributions.


\section{Away from dual-unitarity}
\label{sec:nondu}

\begin{figure}
    \centering
    \includegraphics[width=0.66\columnwidth]{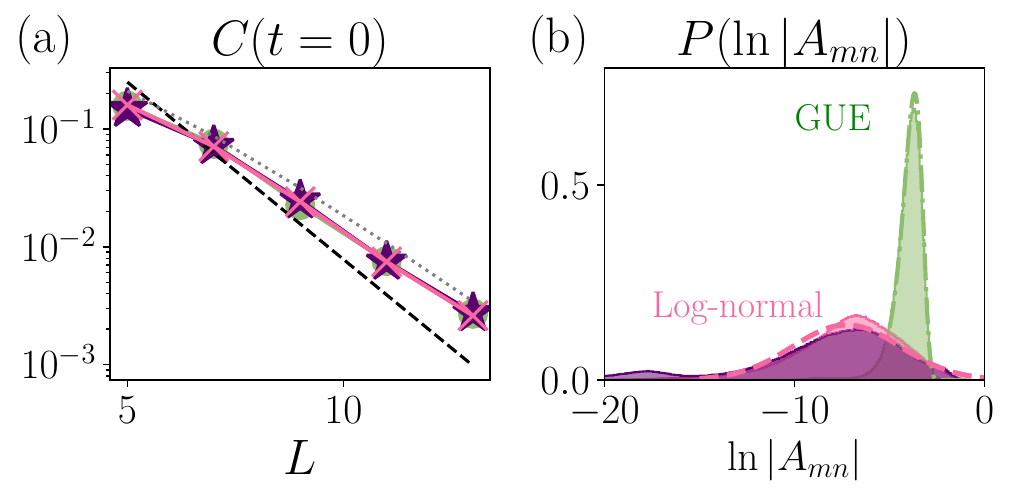}
    \caption{
    (a) Crossing partitions as a function of $L$ for different Trotterization steps, namely $\tau = \pi/4$ (green), $\tau = \pi/4 - 10^{-4}$ (purple, a slight perturbation away from dual-unitarity) and $\tau = \pi/8$ (pink). 
    In all scenarios we observe a quasi-exponential scaling of $L^{3/2}/2^L$, as seen in the gray dotted line. The expected scaling of $1/2^L$ for fully chaotic systems is illustrated by the black dashed line. 
    (b) Corresponding distribution of matrix elements for $L = 13$, omitting zeroes.
    Green (pink) dot-dashed (dashed) line illustrate the fit to a GUE (Log-normal) distribution. 
    }
    \label{fig:crossing_tau}
\end{figure}

We next consider the crossing diagrams for the integrable regime away from the dual-unitary limit $\tau = \pi/4$. 
In this regime the conserved charges no longer consist of solitons, leading to a breakdown of the decomposition in $L \times L$ blocks and eigenstates that no longer satisfy the selection rules used in arriving at Eq.~\eqref{eq:crossing_soliton_basis}.
We numerically compute the scaling of the crossing diagrams for $A = Z_x$ for different values of the Trotter step $\tau$ and consider the corresponding distribution of matrix elements to understand its effect on the crossing and factorization.
As illustrated in Fig.~\ref{fig:crossing_tau}, the scaling is qualitatively the same in all of these scenarios, returning the quasi-exponential scaling of $\sim L^{\frac{3}{2}}/2^L$.
Thus, such a scaling of the crossing partitions seems to be robust even beyond the DU point.
This result indicates that, at least for the local observable considered here, only the most local conservation law (the magnetization) matters and conserved quantities whose densities have larger support are essentially irrelevant. From the point of view of local observables supported on a single site, full ETH does not distinguish between integrable models and ergodic models with only a single and similarly local conservation law.

\begin{figure}
    \centering
    \includegraphics[width=0.66\columnwidth]{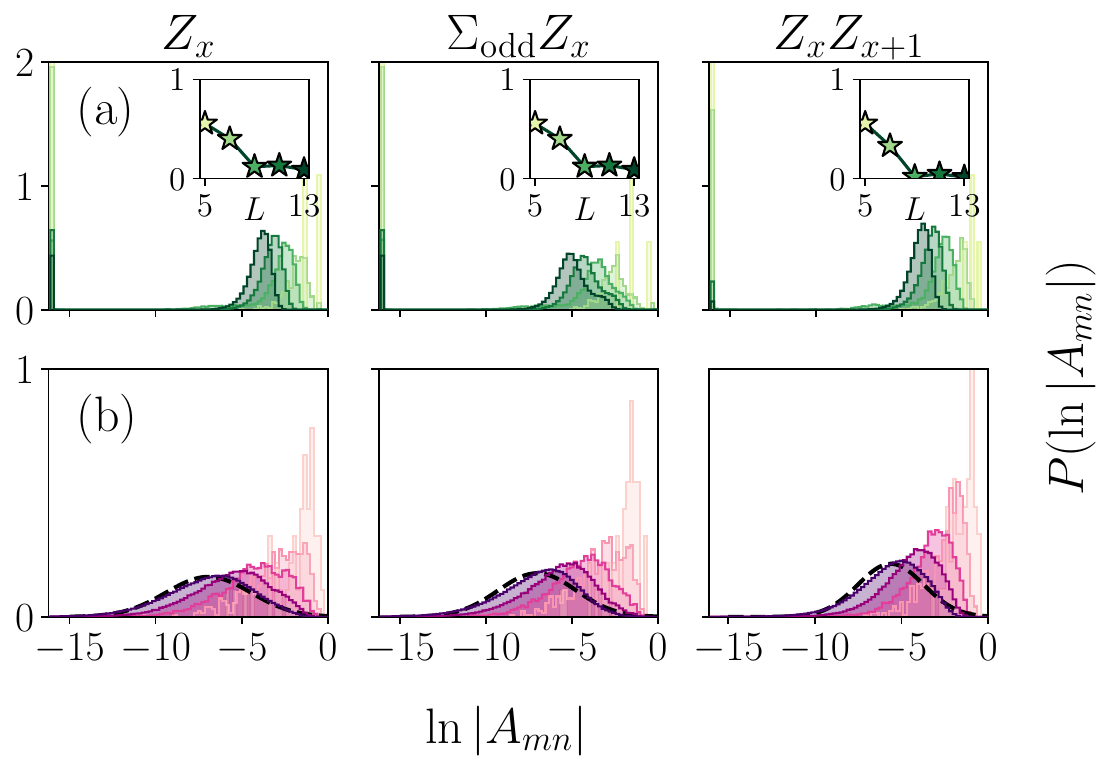}
    \caption{
    Distribution $\ln |A_{ij}| + \mu $ of the logarithm of matrix elements following a projection onto the largest magnetization sector.
    We include a regularization factor $\mu = 10^{-8}$ in order to visualize elements which would be identically zero otherwise. 
    In each of the columns we plot the distribution for (i) the operator $Z$ at the middle of the lattice, (ii) an extensive sum of $Z$ at odd sites, (iii) a $Z \otimes Z$ nearest-neighbor interaction at the middle of the lattice.
    (a) We first consider results for the dual-unitary circuit with $J_z = 1/\sqrt{10}$ and $\tau = \pi/4$.
    (b) We tune the model to the integrable regime away from dual-unitarity by setting $\tau = \pi/3$ while fixing $J_z = 1/\sqrt{10}$.
    The elements are approximately log-normally distributed.
    The lack of dual-unitarity leads to the loss of solitons and extensive conservations laws which follow, as observed in panel (a). Darker colors represent increasingly larger $L$, up to $L = 13$.
    }
    \label{fig:compare_du_nondu}
\end{figure}

This same scaling is however underpinned by a different behavior of the matrix elements, as shown in Fig.~\ref{fig:crossing_tau}.
We consider the distribution of matrix elements in more detail in Fig.~\ref{fig:compare_du_nondu}, comparing different observables at the dual-unitary point and away from the dual-unitary point.
For concreteness, we consider a single-site operator $Z_x$, a sum of local operators $\sum_{x\,\textrm{odd}}Z_x$, and a two-site operator $Z_x Z_{x+1}$.
In row~(a) we illustrate the dual-unitary case, with eigenstates chosen as the random superpositions of the states from Eq.~\eqref{eq:eigenstates}.  Since the eigenstates away from the dual-unitary point are similarly expected to be large linear combinations of such states, this allows for a more direct comparison.
In order to enable visualization, we truncate matrix elements which are identically zero due to the existence of solitons in the model at the dual-unitary point.
In the inset, we show the fraction of non-zero matrix elements, which can be seen to vanish in the limit $L \to \infty$, consistent with the results from the previous section.
In row~(b) we repeat the procedure but away from the dual-unitary point by taking a Trotter step $\tau \neq \pi/4$.
In this case, the model once again displays integrability, but by moving away from dual-unitarity we lose the soliton-based conservation laws present in the DU point.
This difference is evident in Fig.~\ref{fig:crossing_tau}~(b), where even a small perturbation immediately changes the profile of the underlying distribution of matrix elements, even though this change is not detectable in the crossing contributions of Fig.~\ref{fig:crossing_tau}~(a).

The distribution of matrix elements now follows a skewed log-normal distribution, as highlighted with the black dashed line in the last row.
Our result for Floquet circuits thus seem to be consistent with the investigations in Refs.~\cite{leblondEntanglementMatrixElements2019, leblondEigenstateThermalizationObservables2020}, where the authors found this distribution for interacting integrable Hamiltonians (Ref.~\cite{essler_statistics_2024} conversely observed a Fr\'echet distribution in an integrable Lieb-Liniger model).

Overall, in the same way that random matrices whose matrix are randomly but not Gaussian distributed still give rise to free independence (see also Appendix~\ref{app:rmt_crossing}), the different distribution of off-diagonal matrix elements in interacting integrable systems as compare to chaotic systems still gives rise to the suppression necessary for full ETH.
From these numerical investigations, the suppression of crossing diagrams appears to be agnostic to the specific form of the underlying probability distribution of the off-diagonal matrix elements.

\section{Away from integrability}
\label{sec:nonint}

We now consider how these results are modified away from integrability. We choose unitary gates of the form
\begin{equation}\label{eq:su4_parametrization}
    V 
    = U (v_- \otimes v_+)
\end{equation}
where $U$ is the integrable gate of Eq.~\eqref{eq:XXZ_parametrization} and the integrability-breaking terms are of the form
\begin{equation}
    v_{\pm} = \exp(i \epsilon G_{\pm}).
\end{equation}
with $\epsilon \geq 0$ a real parameter used to tune the strength of the on-site disorders and $G_\pm$ are random $2 \times 2$ matrix sampled from the Gaussian Unitary Ensemble (GUE). They are chosen once and kept fixed for all computations.

Moreover, note that our analysis so far was restricted to the time domain.
We mainly focused on the time-zero diagrams (setting $t=0$), since these suffice to characterize the factorization and the suppression of crossing contributions in the model.
For a more detailed picture we now consider how the crossings behave in terms of the quasi-energy differences $\omega$, allowing us to address full ETH in a more refined manner.
More concretely, we now consider the frequency-resolved crossing partition which, in analogy to Eq.~\eqref{eq:crossing}, is defined as:
\begin{equation}\label{eq:crossing_freq}
    C(\omega)
    =
    \frac{1}{D}
    \sum_{i \neq j}
    \delta_\nu(\omega - \omega_{ij})
    |A_{ij}|^4,
\end{equation}
The time-zero results can be recovered by integrating $C(\omega)$ over all frequencies.

\begin{figure*}
    \centering
    \includegraphics[width=\textwidth]{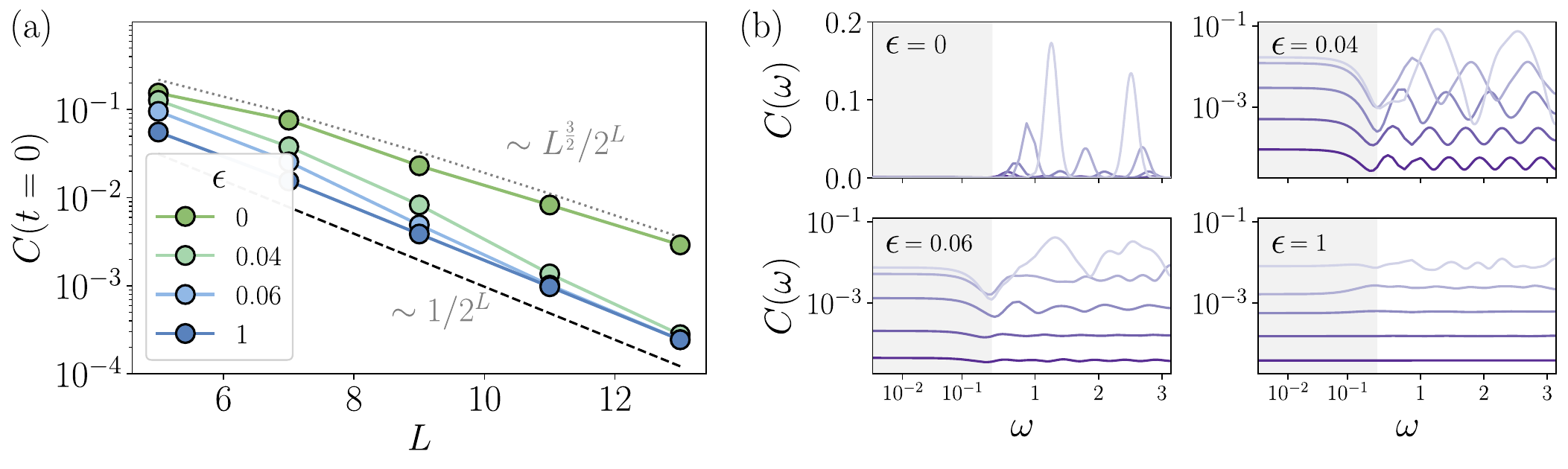}
    \caption{
    Contribution of the crossing diagrams across the parameter space.
    (a) We interpolate from the ergodic regime with on-site disorder with $\epsilon > 0$ to the interacting integrable point at $\epsilon = 0$. We fix $J_z = 1/\sqrt{10}$.
    (b) Crossing contributions as a function of frequency $\omega$. We plot the crossing contributions of panel (a) in the frequency domain. Darker colors represent increasingly larger $L$, up to $L = 13$. Since for $\epsilon=0$ the crossing contribution vanishes identically away from certain frequencies, these results are shown using a linear rather than a logarithmic scale.
    }
    \label{fig:crossing_scaling}
\end{figure*}

Basic results are shown in Fig.~\ref{fig:crossing_scaling}.
In panel~(a) we plot the crossing contributions as a function of system size $L$ for different values of the integrability-breaking parameter $\epsilon$. 
For $\epsilon > 0$, the presence of the on-site unitaries makes the model chaotic, and away from the integrable point we find that the crossing diagrams are suppressed with $1/2^L$, even if it this scaling only becomes apparent for larger systems sizes the closer we are to integrability -- consistent with general results for ergodic models~\cite{pappalardiGeneralEigenstateThermalization2025}.
%
As described in the previous paragraph, in panel~(b) we consider the properties of the model under the same parameters, but in the frequency domain.
Here, $ \delta_\nu(\omega)$ is numerically implemented as a periodized Gaussian with mean zero and standard deviation $\nu$, which needs to be chosen large enough in order to provide numerically consistent results.
We show the plot in a half-log half-linear scale in order to distinguishing between low and high-frequency behavior.
In the integrable limit the factorization follows a Dirac-comb shape, where the number of frequencies increases as the system size is increased. 
This shape directly reflects the selection rules due to the solitons in the dual-unitary integrable limit. 
As the integrability-breaking parameter $\epsilon$ is increased for fixed system size or, equivalently, as the system size is increased for fixed $\epsilon$, this frequency-dependence is lost and the factorization becomes frequency-independent.
In terms of how this relates to the time-domain, note that the OTOC-dynamics is periodic, showing revivals at time-steps are are multiples of the system size, that is, at $t = nL$ with $n \in \mathbb{N}$ (as follows from the fact that $\mathcal{S}^L = \boldsymbol{1}$ and the circuit acts upon the solitons as a swap circuit).
In the frequency domain, the ETH diagrams exhibit delta peaks at frequencies $\omega = 2 \pi n /L$ with $n \in \mathbb{N}$. 
This limiting behavior is consistent with the peaks we see here.

At the integrable point the suppression depends strongly on the choice of frequencies due to this Dirac-comb shape, since the frequency-resolved crossing contribution is only nonzero at a finite number of frequencies set by the eigenphases and new frequencies get introduced at larger values of $L$. 
At the integrable point, we hence only consider a very few specific frequencies where these peaks arise.
For small $\epsilon$ these peaks remain, but get broadened due to the integrability-breaking terms. For sufficiently large $\epsilon$ these peaks are smoothened out and we recover the exponential suppression at all frequencies,
as we can observe at the lower-right panel in Fig.~\ref{fig:crossing_scaling}~(b).

\section{Discussion and conclusion}
\label{sec:conclusion}

The aim of this paper was to investigate which aspects of the full eigenstate thermalization hypothesis still hold in the absence of ergodicity, focusing on interacting integrable quantum circuits. 
Full ETH presents a systematic decomposition of $n$-point correlation functions, including the out-of-time-order correlator, in terms of thermal free cumulants. 
While the conservation laws in integrable systems prevent these free cumulants from being thermal, i.e. only depending on the energy, we here showed how the decomposition in free cumulants still applies even in non-ergodic systems.
The emergence of free cumulants is underpinned by the factorization of cactus (non-crossing) terms and the vanishing of non-cactus (crossing) terms in the expansion of $n$-point correlation functions. 
These properties follow from entropic arguments which break down for noninteracting integrable, i.e. free, models, whereas these can be expected to still apply for interacting integrable models.
We presented an analytical proof that these properties indeed hold in the dual-unitary XXZ model, a particular interacting integrable model, by explicitly deriving the crossing diagrams and showing how they vanish in the thermodynamic limit.
Dual-unitarity constrains the conservation laws of integrability to follow from solitons, and we here used these solitons to derive and fully characterize the eigenstates of this model.
This diagonalization procedure carries over to general dual-unitary circuits which exhibit left and right moving single-site solitons.
The resulting decomposition in cactus and non-cactus diagrams is basis-dependent whenever degeneracies are present, as is the case in this model, which was highlighted by showing how the crossing diagrams can be suppressed either polynomially or exponentially depending on the choice of basis.
One possible way of breaking this degeneracy is by choosing the phases in the dual-unitary circuit to be site-dependent (quenched disorder), which preserves the soliton construction of eigenstates and leads to distinct phases for distinct orbits. While choosing the phases to be site- and time-dependent, i.e. disordered in space and time, preserves the soliton dynamics, there is no longer a meaningful notion of eigenstates and ETH.
To the best of our knowledge, this is the first analytic result on the applicability of free cumulants in the context of full ETH.
We note that dual-unitary circuits mappable to free fermions support multi-site solitons~\cite{holden-dyeFundamentalChargesDualunitary2023}, where the free-fermionic structure however leads to noninteracting dynamics as opposed to the interacting models considered in this work.
The derived exponential suppression was observed in more general interacting integrable models, away from the dual-unitary point, where it now follows from the log-normal distribution of off-diagonal matrix elements.
Moving away from integrability, we observed how the nontrivial frequency-dependence of the crossing diagrams gets washed out as either the integrability-breaking perturbation or system size increases, quickly returning the exponential suppression expected in ergodic systems.

The applicability of (aspects of) full ETH motivates additional studies on the characterization of free probability and the associated time scales for emergent freeness both at the integrable point and for small integrability-breaking perturbations. 
The low-frequency behavior of the second free cumulant presents a sensitive probe for quantum chaos and ergodicity~\cite{pandey_adiabatic_2020,brenesLowfrequencyBehaviorOffdiagonal2020}, and it remains an open question if similar behavior can be observed in higher cumulants.
The dual-unitary XXZ model has recently gained interest since it presents an interacting integrable model in which various quantities that are inaccessible in `generic' integrable systems can be exactly characterized. Dual-unitarity guarantees that correlation functions remain tractable, where this model relaxes to a generalized Gibbs ensemble~\cite{bertiniExactCorrelationFunctions2019,claeys_ergodic_2021}, but the specific combination of dual-unitarity and integrability has allowed for various exact results that cannot be obtained from dual-unitarity alone: This model appeared as a model that is scrambling but not chaotic in the study of local operator entanglement~\cite{bertini_operator_2020,dowling2023scrambling} and its magic and long-range stabilizer R\'enyi entropies can be exactly characterized~\cite{montana_lopez_exact_2024}. The information transmission in such models exhibits dynamical signatures of the propagation and scattering of quasiparticles~\cite{ramppDualUnitarityGeneric2023}, as does its temporal entanglement~\cite{giudice_temporal_2022}, and it is a natural follow-up to relate these results to the exact eigenstates presented in this work.

{
\setlength{\arrayrulewidth}{0.35mm}
\renewcommand{\arraystretch}{1.5}

\begin{table}[h!]
    \centering
    \begin{tabular}{|c|c|cc|}
        \multicolumn{4}{c}{ \normalfont\Large\sffamily \color{quantumviolet} Summary of results} \\[0.5em]
        \hline
        \rule{0pt}{1.5em} 
        {\sffamily \color{quantumgray}  Circuit Type} &  {\sffamily \color{quantumgray} \makecell{Relevant \\ Symmetry}}   & {\sffamily \color{quantumgray} 
 Crossing suppression} & 
        \\[0.2em]
        \hline
        \rowcolor{quantumviolet!10}
        \rule{0pt}{1.5em} 
        & & $C \sim 1/L$ & \makecell{in soliton basis [Eq.~\eqref{eq:eigenstates}]} 
        \\[0.3em]
        \rowcolor{quantumviolet!10}
        \multirow{-2}{*}{\makecell{Dual-unitary, \\[0.3em] interacting and integrable\\[0.3em] [Sec.~\ref{sec:DU_circuits}, Fig.~\ref{fig:crossing_analytical} and~\ref{fig:crossing_fac_analytical_observable}]}} & \multirow{-2}{*}{\makecell{$2^L/L$ soliton- \\[0.3em]based charges}} & $C \sim L^{3/2}/2^L$ & \makecell{in generic basis [App.~\ref{app:random_basis}]} 
        \\[1em]  
        \rule{0pt}{2em} 
        \makecell{Interacting and integrable \\[0.3em] [Sec.~\ref{sec:nondu}, Figs.~\ref{fig:crossing_tau} and~\ref{fig:compare_du_nondu}]} & Magnetization & $C \sim L^{3/2}/2^L$ & --
        \\[1em]   
        \rowcolor{quantumviolet!10}
        \rule{0pt}{1.5em} 
        Chaotic [Sec.~\ref{sec:nonint}, Fig.~\ref{fig:crossing_scaling}] & None & $C \sim 1/2^L$ & --
        \\[0.5em] 
        \hline
    \end{tabular}
    \caption{
    We summarize the three different scenarios we consider. 
    \textbf{(i)} First, the dual-unitary XXZ circuit [parametrized by the gate~\eqref{eq:XXZ_parametrization} with $\tau = \pi/4$], which displays an exponential number of charges~[per Eq.~\eqref{eq:charge_bistring}].
    In the soliton basis we obtain the (analytical) polynomial suppression of crossing diagrams~[Eq.~\eqref{eq:crossing}] we observe in Eq.~\eqref{eq:crossing_analytical}.
    In a generic basis, which can be constructed as a superposition of eigenstates in a degenerate subspace, we observe a quasi-exponential scaling following random matrix theory calculations in App.~\ref{app:random_basis}.
    These results are show in Fig.~\ref{fig:crossing_analytical} and~\ref{fig:crossing_fac_analytical_observable}.
    \textbf{(ii)} 
    We strip-away dual-unitarity from the circuit [using Eq.~\eqref{eq:XXZ_parametrization} with $\tau \neq \pi/4$], resulting in an integrable circuit with an extensive number of conservation laws, where eigenstates with a fixed magnetization again lead to a quasi-exponential scaling.
    These results are show in Figs.~\ref{fig:crossing_tau} and~\ref{fig:compare_du_nondu}.
    \textbf{(iii)} 
    We consider a chaotic circuit with no conserved quantities [parametrized as in Eq.~\eqref{eq:su4_parametrization} to break integrability], which leads to the known scaling of $1/2^L$ for the suppression of crossing diagrams, as shown in Fig.~\ref{fig:crossing_scaling}.
    }
    \label{tab:placeholder}
\end{table}
}

The numerical routines used in this manuscript and a minimal example are available at~\cite{alvesAlvesgabrielDual_unitary_ethFirst2025}.

\appendix

\section*{Acknowledgements}
We acknowledge useful discussions with S. Pappalardi, M.A. Rampp, and E. Vallini and support by the Max Planck Society and the Max Planck Computing and Data Facility.
F.F. acknowledges support from the European Union's Horizon Europe program under the Marie Sk{\l}odowska Curie Action GETQuantum (Grant No. 101146632).


\appendix


\section{Further details on the exact diagonalization of the interacting integrable DU circuit}\label{app:diagonalization}

\subsection{Characterization of subsectors}\label{sec:circuit_sectors}
\label{app:charge_sector}

In this Appendix we characterize the smallest soliton subsectors in the dual-unitary integrable model.
We first show that all subsectors are comprised of a representative state and all the permutations thereof.
In order to do so, let us assume that we have a charge $Q_{(a)}$ [Eq.~\eqref{eq:charge_bistring}] and a state $\ket{a}$ constructed from the corresponding bitstring $a$ [Eq.~\eqref{eq:eigenstates}], as well as two states $\ket{m}$ and $\ket{n}$ which by assumption belong to different sectors, i.e. they are not permutations of each other. 
We should then be able to associate these sectors (and their bitstrings) to two distinct charges $Q_{(m)}$ and $Q_{(n)}$. 
That is, $\ket{m} \neq \mathcal{S}^t \ket{n}$, $\forall t \in \mathbb{Z}$ and $Q_{(m)} \neq Q_{(n)}$.
By setting up the problem like this, we have the following eigenvalues/quantum numbers:
\begin{equation}
    Q_{(a)} | m \rangle = \lambda^{a}_m | m \rangle, \qquad
    Q_{(a)} | n \rangle = \lambda^{a}_n | n \rangle.
\end{equation}
Now, we shall also assume that $\lambda^{a}_m = \lambda^{a}_n = \lambda^{a}$ for \emph{all} bitstrings $a$ with $|a|$ non-zero bits. 
That is, we are assuming that the states $\ket{m}$ and $\ket{n}$ have exactly the same quantum numbers for \emph{all} charges $Q_{(a)}$, even though they are \emph{not} permutations of each other. Naturally, we also assume that $|m| = |n|$.

Due to the soliton-state duality in Eq.~\eqref{eq:soliton_state_duality}, the charge $Q_{(m)}$ acts on $| a \rangle $ as
\begin{equation}
    Q_{(m)} | a \rangle = \lambda^{(a)} | a \rangle =
    (-1)^{|m| - |a|} 
    \lambda^{a}_m | a \rangle.
\end{equation}
But, by assumption, we know that $\lambda^a_m = \lambda^{a}_n$. 
Moreover, we can once again use the dual property $\lambda^{a}_n = (-1)^{|a| - |n|}  \lambda^{n}_a$, to rewrite the expression above in terms of $n$ instead:
\begin{equation}
    Q_{(m)} | a \rangle = \lambda^a_m | a \rangle =
    (-1)^{|m| - |a|} 
    (-1)^{|a| - |n|} 
    \lambda^n_a | a \rangle.
\end{equation}
This is still true for all $a$. Using that $|m| = |n|$ by assumption, the prefactor above cancels out and we get
\begin{equation}
    Q_{(m)} | a \rangle =
    Q_{(n)} | a \rangle,
    \quad
    \forall \: a.
\end{equation}
This result violates our initial assumption that $(m)$ and $(n)$ are unrelated by permutations. 
We conclude that, if all quantum numbers coincide, the states $m$ and $n$ are necessarily permutations of each other, i.e. $\ket{n} = \mathcal{S}^t \ket{m}$ for some $t \in \{0, ..., L - 1\}$.

Any state hence corresponds to an orbit that is maximally of length $L$.
In order to obtain the possible lengths of the orbits, we note that the group of translations is isomorphic to the cyclic group with $L$ elements, has order $L$ and acts on the set of computational basis states as a translation (see also the discussion in Appendix~\ref{app:fkm_algorithm}).
By the orbit-stabilizer theorem~\cite{carrellTheoremsGroupTheory2017} the length of the orbit has to divide the order $L$ of the full translation group. For $L$ prime, we find that orbits are thus either of length $L$ or $1$, with the latter corresponding solely to the fully polarized states.
In this case all sectors are hence either $1 \times 1$ or $L \times L$ in dimensionality.

Circuits with a non-prime length have slightly more complicated combinatorics, with orbits which can have the same length as any of the divisors of $L$. 
A concrete example for $L=9$ is given by the bitstring $001100001$, and the corresponding charge follows from an orbit of size $3$ as
$Q_{001100001} = 
\scalebox{0.75}{\centering \leg \leg \soliton \soliton \leg \leg \leg \leg \soliton} +
\scalebox{0.75}{\centering \leg \soliton \leg \leg \soliton \leg \leg \soliton \leg} +
\scalebox{0.75}{\centering \soliton \leg \leg \leg \leg \soliton \soliton \leg \leg}. 
$

This result also allows us to make some statements about the dimensionality of the system. 
First, since all sectors with the exception of the fully polarized states have dimension $L$, we need a total of $(2^L - 2)/L \sim 2^L/L$ quantum numbers to fully describe the system.
Due to Fermat's little theorem, this number is indeed an integer.
Similarly, if we restrict ourselves to fixed magnetization sectors, the number of unique representative states, or equivalently, of unique $|m|$-body charges is given by:
\begin{equation}\label{eq:dimensionality_solitons}
    d_{L, |m|}
    =
    \# \text{ distinct $|m|$-body charges}
    =
    \frac{1}{L}
    {L \choose |m|}.
\end{equation}
In particular we have that the number of two-body charges is given by $d_{L, 2} = (L - 1)/2$.
Moreover, by using Stirling's approximation, we recover the scaling
\begin{equation}
    d_{\max}
    =
    d_{L, (L + 1)/2}
    \sim
    O\left(\frac{2^L}{L^{3/2}}\right)
\end{equation}
for the number of representative states in the largest magnetization sector (or number of $(L \pm 1)/2$-body charges).

\subsection{Scattering matrix}\label{app:scattering_matrix}

\begin{figure}
    \centering
    \includegraphics[width=0.5\columnwidth]{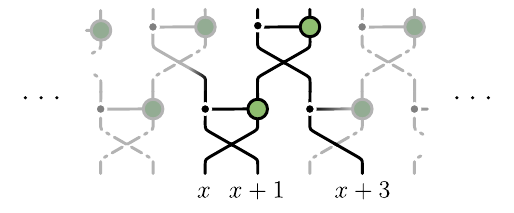}
    \caption{
    In the bulk, the spins on sites $x + 1$ and $x + 3$ act as a control on the right on the spin in the right-moving site $x$. 
    }
    \label{fig:scattering_picture}
\end{figure}

In this Appendix we detail the calculation of the partial scattering phases.
In order to compute the individual phases $\phi_s$, note that the right-moving spin in the site $x$ acquires a phase depending on the state in sites $x + 1$ and/or $x + 3$, acting as control qubits, as shown in Fig.~\ref{fig:scattering_picture}.
At the boundary, the spin in the site $L - 2$ scatters with the spins in the sites $L - 1$ and $ L $.
This means that a state $m$ acquires a phase $\phi_s = \phi \, (m| V |m)$, where $V$ is a $L \times L$ scattering matrix:
\begin{equation}\label{eq:unitary_sector}
    V
    =
    \begin{matrix}
    \begin{pmatrix}
    0 & 1 & 0 & 1 & 0 & 0 &  \hdots & 0 & 0 \\ 
    0 & 0 & 0 & 0 & 0 & 0 &  \hdots & 0 & 0 \\ 
    0 & 0 & 0 & 1 & 0 & 1 &  \hdots & 0 & 0 \\
    \vdots & \vdots & \vdots & \vdots & \vdots & \vdots & \ddots & \vdots & \vdots \\
    0 & 0 & 0 & 0 & 0 & 0 &  \hdots & 1 & 1 \\ 
    0 & 0 & 0 & 0 & 0 & 0 &  \hdots & 0 & 0 \\ 
    0 & 0 & 0 & 0 & 0 & 0 &  \hdots & 0 & 0 
    \end{pmatrix}
    \end{matrix}\,.  
\end{equation}

In this convention we have ordered the permutations from top to bottom.


\section{Distribution of charges}\label{app:charge_distribution}

As we argue in the main text, the crossing diagrams can be expressed in terms of the distribution of one- and two-body charges. 
In this Appendix, we consider the statistics of these charges.
In particular, we consider the fourth moment of the magnetization, which is the unique charge constructed from the single-body solitons, and the second moment for the distribution of two-body charges.
For the former we note that the eigenvalues can assume the values $\lambda_{a, |a| = 1}^m = 2|m| - L = L, L - 2, ..., 1, -1, ..., - L + 2, L$, which are binomially distributed.
We can write down the probability mass function (pmf) for the magnetization/one-body charge as:
\begin{equation}\label{eq:binomial_magnetization}
    f_{|a| = 1}(\lambda; L)
    =
    \frac{1}{2^L}
    {L \choose \frac{\lambda + L}{2}},
    \quad
    \lambda 
    =
    L, L - 2, ...,  -L
    .
\end{equation}

For the two-body charges we note that each term in the two-body charge consists of a parity check between the spins in two sites $x$ and $y$, as in the examples:
\[
\begin{split}
    & (\hdots \: \source{\underset{x}{\soliton}} \: \hdots \: \underset{y}{\soliton} \: \hdots )
    \ket{\hdots \source{0} \hdots \target{0} \hdots} = + 1 \drawarrows \\
    & (\hdots \: \underset{x}{\soliton} \: \hdots \: \underset{y}{\soliton} \: \hdots )
    \ket{\hdots \source{0} \hdots \target{1} \hdots} = - 1. \drawarrows
\end{split}
\]
For all spins up (or down), there are no mismatches and we have $\lambda_{a, |a| = 2}^m = L$. 
Similarly, if we have two mismatches (which can be obtained by e.g. flipping a single spin in the fully polarized state), we get $\lambda_{a, |a| = 1}^m = L - 4$. 
More generally, if we have $M$ mismatches among the $L$ terms appearing in a two-body charge, the associated quantum number will be $\lambda_{a, |a| = 2}^m = (L - M) - M = L - 2M$, as represented in:
\vspace{10pt}
\[
\begin{split}
    L - M
    & \begin{cases}
    & \hspace{-8pt} \soliton \hdots \: \soliton \hdots \: \leg \: \hdots \: \leg \: \hdots \: \leg
    \ket{ \source{1} \hdots \target{1} \hdots 1 \hdots 0 \hdots 1} + \drawarrows \vspace{3pt} \\
    & \hspace{-8pt} \soliton \hdots \: \leg \hdots \: \soliton \: \hdots \: \leg \: \hdots \: \leg
    \ket{ \source{1} \hdots 1 \hdots \target{1} \hdots 0 \hdots 1} + \drawarrows \vspace{3pt} \\
    &  \hspace{-8pt} \leg \hdots\: \soliton \hdots \: \soliton \: \hdots \: \leg \: \hdots \: \leg
    \ket{ 1 \hdots \source{1} \hdots \target{1} \hdots 0 \hdots 1} + \drawarrows \vspace{3pt} \\
    \end{cases}
    \\
    - M
    & \begin{cases}
    & \hspace{-8pt} \leg \hdots \: \leg \hdots \: \soliton \: \hdots \: \soliton \: \hdots \: \leg
    \ket{ 1 \hdots 1 \hdots \source{1} \hdots \target{0} \hdots 1} + \drawarrows \vspace{3pt}\\
    &  \hspace{-8pt} \leg \hdots \:\leg \hdots \: \leg \: \hdots \: \soliton \: \hdots \: \soliton
    \ket{ 1 \hdots 1 \hdots 1 \hdots \source{0} \hdots \target{1}}. \drawarrows
    \end{cases}
\end{split}
\]
\vspace{10pt}

Due do the fact that each site contributes to two different terms in the charge, we can only have an even number of mismatches.
The values that the two-body quantum numbers can assume follow as $L, L - 4, L - 8, ..., - L + 2$, half as many as the magnetization.
By counting how these mismatches can be distributed in Eq.~\eqref{eq:charge_bistring} we find the correct combinatorics for the two-body charges as $2 {L \choose M}$. 
Note that we have an extra factor of two here, since the two-body charges display inversion symmetry.
This is due to the fact that we can write down these mismatches with either $x$ being a spin up and $y$ being a spin down or vice-versa.
The pmf for the quantum numbers of two-body charges is thus:
\begin{equation}\label{eq:binomial_two_body}
    f_{|a| = 2}(\lambda; L)
    =
    \frac{1}{2^{L - 1}}
    {L \choose \frac{\lambda + L}{2}},
    \quad
    \lambda 
    =
    L, L - 4, ...,  - L + 2.
\end{equation}
Note the different normalization factor from Eq.~\eqref{eq:binomial_magnetization}.
These results recover the densities~\eqref{eq:binomial_magnetization} and~\eqref{eq:binomial_two_body} as shown in Fig.~\ref{fig:charge_distribution} in the main text.

We can thus simply compute both of these momenta from the binomial distribution.
For the magnetization, we get 
$\expec{\lambda_{a, |a| = 1}^4} = 3L^2 - 2L$.
Similarly, for the two-body charges we get $ \expec{\lambda_{a, |a| = 2}^2} = L$, which is coincidentally the same we would get for the magnetization. 


\section{Crossing diagrams for the integrable DU circuit}\label{app:crossing_DU}

Let us compute the crossing contribution in a fixed sector labeled by the representative state $m$, for a local observable $Z_x$ given by a soliton at site $x$:
\begin{align}
    C_m = \frac{1}{L} \sum_{k \neq l}
    \left|\bra{\varepsilon_k^m}Z_x\ket{\varepsilon_l^m}\right|^4.
\end{align}
For ease of notation we denote the canonical basis states of the sector by $r,s,t, u$ and $v$., which are obtained by applying the swap circuit to the representative state $m$ in the computational basis.
That is, $\ket{s} = \ket{\sigma^s(m)}$ and so on.

The matrix elements of $Z_x$ in the canonical basis can be written as
\begin{align}
    \bra{s}Z_x\ket{r} 
    = 
    \delta_{r,s} (-1)^{[\sigma^r (m)]_x} 
    = 
    \delta_{r,s} (-1)^{m_{\sigma^{-r}(x)}}, 
\end{align}
where we define $m_{\sigma^{-r}(x)} = [\sigma^r (m)]_x$. 
Consequently, using $\theta_l - \theta_k = 2\pi(l-k)/L$, the matrix elements in the eigenbasis read
\begin{align}
    \bra{\varepsilon_k^m}Z_x\ket{\varepsilon_l^m} & =  \frac{1}{L}\sum_{r,s}e^{i \left(\theta_l r - \theta_k s  + \Phi_r - \Phi_s \right)}\bra{s}Z_x\ket{r} \\
    & =  \frac{1}{L}\sum_{r}e^{i \left(\theta_l - \theta_k \right) r} (-1)^{m_{\sigma^{-r}(x)}} \\
    & =  \frac{1}{L}\sum_{r}e^{i 2\pi(l-k)r/L - i \pi m_{\sigma^{-r}(x)} }.
    \label{eq:Z_x_matrix_elements_sector}
\end{align}
As we can see, obtaining the matrix elements for $Z$ is particularly simple since the result does not depend on the partial scattering phase.
This expression immediately gives the diagonal matrix elements in terms of the total magnetization $M=\lambda_{e_x}^m$, i.e.,
\begin{align}
\bra{\varepsilon_l^m}Z_x\ket{\varepsilon_l^m}  =  -\frac{\lambda_{e_x}^m}{L} = 1 - \frac{2|m|}{L} \in (-1, 1).
\end{align}
independent from $l$.
Here we denote the standard basis in $\mathbb{Z}_2^L$ by $e_x$.
As the diagonal matrix elements are of order one, one has
\begin{align}\label{eq:diagonal_elements}
    \frac{1}{L} \sum_{l}
    \left|\bra{\varepsilon_l^m}Z_x\ket{\varepsilon_l^m}\right|^4 = \frac{\left(\lambda_{e_x}^m\right)^4}{L^4}  \in (-1, 1). 
\end{align}
The crossing contribution can be written as 
\begin{align}
    C_m = \frac{1}{L} \sum_{k,  l}
    \left| \bra{\varepsilon_k^m}Z_x\ket{\varepsilon_l^m}\right|^4 - \frac{M^4}{L^4}.
\end{align}
Moreover, as $\bra{\varepsilon_k^m}Z_x\ket{\varepsilon_l^m}$ depends only on $(k-l)\mod \: L$ we have
\begin{align}
    \frac{1}{L} \sum_{k,  l}
    \left|\bra{\varepsilon_k^m}Z_x\ket{\varepsilon_l^m}\right|^4 
    = \sum_{l}
    \left|\bra{\varepsilon_0^m}Z_x\ket{\varepsilon_l^m}\right|^4 \nonumber
    = \frac{M^4}{L^4} +  \sum_{l=1}^{L-1}
    \left|\bra{\varepsilon_0^m}Z_x\ket{\varepsilon_l^m}\right|^4.
\end{align}
It is however convenient to keep the $l=0$ term and compute 
\begin{equation}
\begin{split}
    C_m + \frac{M^4}{L^4} & = \sum_{l=0}^{L-1}
    \left|\bra{\varepsilon_0^m}Z_x\ket{\varepsilon_l^m}\right|^4 \\
    & = 
    \frac{1}{L^4}\sum_{l=0}^{L-1}\sum_{r, s, t, u}
    e^{2\pi i l (r + s + t + u)/L}
    \times 
    e^{i \pi \left(m_{\sigma^{-r}(x)}+ m_{\sigma^{-s}(x)} + m_{\sigma^{-t}(x)} + m_{\sigma^{-u}(x)}\right)} \\
    & = 
    \frac{1}{L^4}\sum_{r, s, t, u}e^{i \pi \left(m_{\sigma^{-r}(x)}+ m_{\sigma^{-s}(x)} + m_{\sigma^{-t}(x)} + m_{\sigma^{-u}(x)}\right)} 
    \times \left(\sum_{l=0}^{L-1}e^{2\pi i l (r + s + t + u)/L}\right).
\end{split}
\end{equation}  
The last term yields
\begin{align}
    \sum_{l=0}^{L-1}e^{2\pi i l (r + s + t + u)/L} 
    = 
    L \delta_{r + s, t + u} 
    = 
    L \sum_{v=0}^{L-1}\delta_{r, t + v} \delta_{s+v, u} 
\end{align}
with all $\delta_{a,b}$ being understood $\!\! \! \mod \! L$ and hence
\begin{equation}
\begin{split}
    C + \frac{M^4}{L^4} 
    & = 
    \frac{1}{L^3}\sum_{r, s, t, u, v}e^{i \pi \left(m_{\sigma^{-r}(x)}+ m_{\sigma^{-s}(x)} + m_{\sigma^{-t}(x)} + m_{\sigma^{-u}(x)}\right)}\delta_{r, t + v} \delta_{s+v, u} \\
    & = 
    \frac{1}{L^3}\sum_{s, t, v}e^{i \pi \left(m_{\sigma^{-t-v}(x)}+ m_{\sigma^{-s}(x)} + m_{\sigma^{-t}(x)} + m_{\sigma^{-s-v}(x)}\right)} \\
    & =  
    \frac{1}{L^3}\sum_{v}
    \Bigg[
    \left(
    \sum_s e^{i \pi \left(m_{\sigma^{-s}(x)} + m_{\sigma^{-s-v}(x)}\right)}
    \right)
    \times
    \left(
    \sum_t e^{i \pi \left(m_{\sigma^{-t}(x)} + m_{\sigma^{-t-v}(x)}\right)} \right) 
    \Bigg]\\
    & =  
    \frac{1}{L^3}\sum_{v}
    \Bigg[
    \left(
    \sum_s (-1)^{\left(m_{\sigma^{-s}(x)} + m_{\sigma^{-s-v}(x)}\right)}
    \right)
    \times
    \left(
    \sum_t (-1)^{\left(m_{\sigma^{-t}(x)} + m_{\sigma^{-t-v}(x)}\right)} \right) 
    \Bigg]\\
    &  = 
    \frac{1}{L} + \frac{1}{L^3}\sum_{v=1}^{L-1}
    \Bigg[
    \left(
    \sum_s (-1)^{\left(m_{\sigma^{-s}(x)} + m_{\sigma^{-s-v}(x)}\right)}
    \right)\\
    &\quad \quad \quad \quad \quad \quad \: \: \times
    \left(
    \sum_t (-1)^{\left(m_{\sigma^{-t}(x)} +  m_{\sigma^{-t-v}(x)}\right)} \right)
    \Bigg].
\end{split}
\end{equation}
In the last line we have used that at $v=0$ the sum over $s$ yields $L$ and so does the sum over $t$.
Denoting $a_v = e_x + e_{\sigma^{v}(x)}$ both the sum over $s$ and $t$ yield the eigenvalues $\lambda_{a_v}^m$ for the two-body charge $Q_{a_v}$.
By summing over $v$ we sum twice over the $(L-1)/2$ unique two-body charges, returning Eq.~\eqref{eq:crossing_soliton_basis} from the main text.
We can thus see that knowledge of the one and two-body charges is enough to compute the crossing partitions for $Z_x$.


\section{Diagonal matrix elements of the operator $A$}\label{app:diagonal_of_operator}

For the class operators we discuss here, namely operators of the form $A=Z \cos \theta + X \sin \theta$ with $\theta \in [0, \pi/2]$, we can show how diagonal matrix elements can be disregarded in the computations from Sec.~\ref{sec:FP_ETH}. While for generic operator setting the matrix elements equal to zero corresponds to a highly nonlocal perturbation of the operator, for the observables considered in this work this can be done by subtracting a local operator, as shown in this Appendix.

We found in the previous section that the diagonal entries $\bra{\varepsilon_k^m} Z \ket{\varepsilon_k^m}$ of $Z$ are given by Eq.~\eqref{eq:diagonal_elements} and only depend on $m$. Similarly, we have that $\bra{\varepsilon_k^m} X \ket{\varepsilon_k^m} = 0$ for any state $\ket{\varepsilon_k^m}$ in the soliton basis regardless of the site.
Thus, just as in Eq.~\eqref{eq:diagonal_elements}, we have that
$ \bra{\varepsilon_k^m} A \ket{\varepsilon_k^m} = 1 - 2|m| / L$
for all magnetization $m$ and momentum $k$.
We can therefore consider the shifted operator
\begin{equation}
    \tilde{A}
    =
    A - \cos{\theta} \sum_{m, k} \left(1 - \frac{2|m|}{L}\right)\ketbra{\varepsilon_k^m}{\varepsilon_k^m}
    =
    A - S_z \cos{\theta},
\end{equation}
in all our computations, where $S_z$ is the magnetization operator given by Eq.~\eqref{eq:magnetization_op}.
This way we ensure that we have vanishing diagonal elements $\tilde{A}_{ii} = 0$, which greatly simplifies further computations and makes the decomposition from Eq.~\eqref{eq:ETH_decomposition} onwards valid.
Also note that the off-diagonal elements of $A$ are left unchanged under this transformation.


\section{Crossing diagrams in degenerate eigenspace}
\label{app:random_basis}
Here we present an estimate for the scaling of the crossing diagrams for $Z_x$ in a generic basis of the degenerate eigenspaces of the evolution operator~\eqref{eq:circuit_int}, recovering quasi-exponential scaling as observed in Fig.~\ref{fig:crossing_scaling}.
Such an eigenspace is characterized by the momentum $k$ and the magnetization $M$ and is spanned by the states $|\varepsilon_k^{m} \rangle$ with fixed $k$ and $m$ ranging over all the distinct representative states with magnetization $M$. Denoting the dimension of the subspace by $d=d(k, M)$ and arbitrary labeling the above basis states by $|\tilde{\varepsilon}_i\rangle$ for $1 \leq i \leq d$ we might construct a generic basis 
\begin{equation}\label{eq:rotated_eigenstate}
   \ket{\alpha} = \sum_{i=1}^dW_{\alpha, i} \ket{\tilde{\varepsilon}_i},
\end{equation}
by rotating with a Haar random $d\times d$ unitary $W$.
Using the representation of matrix elements Eq.~\eqref{eq:Z_x_matrix_elements_sector} as well as the fact that $Z_x$ does not couple different sectors we find
\begin{equation}
    \langle  \varepsilon_i | Z_x |\tilde \varepsilon_j \rangle = \frac{M}{L}\delta_{ij} \, .
\end{equation}
In the rotated basis the matrix elements read
\begin{equation}
    \langle \alpha |Z_x | \beta \rangle = \frac{M}{L}\sum_{i=1}^ d W_{\alpha, i}^*W_{\beta, i} \, 
\end{equation}
and these contribute to the crossing term as
\begin{align}
& |\langle \alpha |Z_x|\beta \rangle|^4 \nonumber \\
&\,\,= \frac{M^4}{L^4}\sum_{i_1,\ldots,i_4}W_{\beta, i_1}W_{\beta, i_2}W_{\alpha, i_3}W_{\alpha, i_4}
 W_{\alpha, i_1}^*W_{\alpha, i_2}^*W_{\beta, i_3}^*W_{\beta, i_4}^* \, ,
\end{align}
with $*$ denoting complex conjugation.
To compute the crossing for a generic basis we take the Haar average over $W$ of the above expression.
Assuming large $d\gg 1$ and $\alpha\neq\beta$, to leading order in $1/d$ this yields \cite{collins2006integration}
\begin{equation}\label{eq:expected_crossing_permutation}
\mathbb{E}_W  |\langle \alpha |Z_x|\beta \rangle|^4 = \left(\frac{M}{ d L}\right)^4\sum_{i_1,\ldots,i_4}\sum_{\pi}\prod_{j=1}^4\delta_{i_j,i_{\pi(j)}}
\end{equation}
with $\pi$ running over the permutations $(13)(24)$, $(14)(23)$, $(1423)$, and $(1324)$, denoted here in cycle notation.
Evaluating the sum $\sum_{i_j}\prod_j \delta_{i_j,i_{\pi(j)}}$ yields
a term $\sim d^2$ for the first two permutations and a subleading term $\sim d$ for the last two permutations, as the latter force all the $i_j$ to be equal.
Consequently,  $ |\langle \alpha |Z_x|\beta \rangle|^4  \sim (M/L)^4d^{-2}$ is of order $\sim d^{-2}$ and therefore the sum over $\sim d^2$ basis states $\alpha$ and $\beta$ as given by
\begin{equation}\label{eq:crossing_sum_rmt}
    \sum_{\alpha \neq \beta}  |\langle \alpha |Z_x|\beta \rangle|^4  \sim \left(\frac{M}{L}\right)^4 
\end{equation}
is of order 1.
We obtain the crossing by summing the above result over all $L$ quantum numbers $k$ and all possible magnetizations $M$ as
\begin{align}
     C \sim \frac{1}{D}\sum_{k}\sum_M \left(\frac{M}{L}\right)^4 = \frac{1}{DL^3}\sum_M M^4 \sim \frac{L^2}{D} \, ,
\end{align}
using $\sum_M M^4 \sim L^5$. Fixing $D=2^L$, this yields only a slightly weaker suppression as the estimate based on the largest eigenspaces used in the main text.



\section{Generating representative states}\label{app:fkm_algorithm}

As discussed in the main text, we can diagonalize the dual-unitary XXZ circuit and obtain all the eigenstates through the identification of representative states $m$ which label different equivalence classes.  
They uniquely define the set of $L$ bitstring permutations $\ket{\pi^s(m)}\}_{s = 0, 1, ..., L - 1}$ which constitute a sector. In this Appendix we briefly discuss how such representative states can be efficiently generated, which is a necessary step in any numerical construction of the eigenstates.

A naive implementation for obtaining representative states is costly, since in the worst case scenario one would have to compare a given bitstring with all the other $2^L/L$ representative states.
In order to efficiently generate these states, we resort to well known results in the field of \emph{necklaces}, a nomenclature used in combinatorics literature to denote strings with translational/rotational invariance.
We borrow a central result therein, the so-called FKM algorithm -- owing to Fredricksen, Kessler and Maiorana~\cite{fredricksenNecklacesBeadsColors1978, fredricksenAlgorithmGeneratingNecklaces1986}. 
An accessible introduction and historical overview on the subject can be found in Ref.~\cite{sawadaGeneratingBraceletsConstant2001b}. 

In this section we modify our notation to a more convenient form in order to discuss the aforementioned literature.
The permutation $\sigma$ defined in Eq.~\eqref{eq:permutation_def_cycle} can alternatively be defined in cycle notation as:
\begin{equation}\label{eq:permutation_def_cycle}
    \sigma = (1\; 3\; 5\; ...\; L\;\; L-1\; ...\; 4\; 2).
\end{equation}
By doing so, we can get a simpler representation of the bitstrings and their permutations.
With a slight abuse of notation, we write the bitstrings in a similar notation as
\begin{equation}\label{eq:cycle_soliton}
    (m_1 ... m_L) 
    \leftrightarrow 
    \bigotimes_{i = 1}^{L}
    Z^{m_i}_{\sigma[i]},
\end{equation}
an alternative notation to Eq.~\eqref{eq:soliton_bitstring}, but ``translated" to cycle notation. 
We can make the definition above more transparent through graphical notation. For instance:
\[
\begin{split}
    (11000) &\leftrightarrow \soliton \leg \soliton \leg \leg \\
    (01100) &\leftrightarrow \leg \leg \soliton \leg \soliton \\
    (00110) &\leftrightarrow \leg \leg \leg \soliton \soliton \\
    (00011) &\leftrightarrow \leg \soliton \leg \soliton \leg \\
    (10001) &\leftrightarrow \soliton \soliton \leg \leg \leg. \\
\end{split}
\]

With the example above, this notation makes clear how to obtain all the states in a sector from its representative: we can simply translate the corresponding bitstring $L$ times. 
This notation is convenient because this is precisely how the problem of necklace generation is posed in the literature.
Additionally, as long as we are still dealing with permutations, the formulation in terms of the cycle notation is independent of the circuit architecture, such that we could use this formalism and the FKM algorithm to generate representative states also for different configurations, including periodic boundary conditions.

\begin{table}
    \centering
    \begin{tabular}{  c  c  c}
    \vspace{5pt}
     (00000) & & \leg \leg \leg \leg \leg \\ \vspace{5pt}
     (00001) & & \leg \leg \leg \soliton \leg\\ \vspace{5pt}
     (00011) & & \leg \soliton \leg \soliton \leg  \\ \vspace{5pt}
     (00101) & $ \leftrightarrow$ & \leg \soliton \leg \leg \soliton\\ \vspace{5pt}
     (00111) & & \leg \soliton \leg \soliton \soliton \\ \vspace{5pt}
     (01011) & & \leg \leg \soliton \soliton \soliton \\ \vspace{5pt}
     (01111) & & \leg \soliton \soliton \soliton \soliton \\ \vspace{5pt}
     (11111) & &  \soliton \soliton \soliton \soliton \soliton 
    \end{tabular}
    \caption{Representative bitstrings $\mathcal{R}_5$ and the corresponding solitons.}
    \label{table:rep_bitstrings}
\end{table}

We follow the implementation of the FKM algorithm of Ref.~\cite{ruskeyGeneratingNecklaces1992a}. 
The algorithm is very efficient and runs on constant amortized time.
Starting with the bitstring $0^L \equiv \underbrace{0 \hdots 0}_{L \text{ times}}$, the algorithm sequentially generate the representative states in lexicographical order, until the last string $1^L$ is reached.
The pseudocode, simplified for bitstrings, is presented in Algorithm~\ref{alg:FKM-algorithm}. We denote the set of the set of representative bitstrings which label the equivalence classes by $\mathcal{R}_L$.
\algnewcommand{\Initialize}[1]{%
  \State \textbf{Initialize:}
  \Statex \hspace*{\algorithmicindent}\parbox[t]{.8\linewidth}{\raggedright #1}
}

\begin{algorithm}[H]
   \caption{The FKM algorithm for binary strings.}
   \label{alg:FKM-algorithm}
   \begin{algorithmic}[1]
        \Procedure{FKM}{L}
        \State \textbf{Initialize:}
        \State \hspace{13pt} $ a \gets 0^L$
        \State \hspace{13pt}  $\mathcal{R}_L \gets [a]$ \Comment{List of representative strings}
        \State \hspace{16pt}$i \gets L$
        \While {$i > 0$} \Comment{Stops at $1^L$}
            \State $a[i] \gets a[i] + 1$
            \For{$j \gets 1, ..., L - i$} $a[j + i] \gets a[j]$\EndFor
            \If {$n \mod i = 0$}  $\text{append $a$ to } \mathcal{R}_L$ \EndIf
            \State $i \gets L$
            \State \textbf{repeat} {$i \gets i - 1$} \textbf{until} {$a[i] = 1$}
        \EndWhile
        \State \textbf{return} $\mathcal{R}_L$
        \EndProcedure
   \end{algorithmic}
\end{algorithm}

For $L = 5$ one has $(2^5 - 2)/5 + 2 = 8$ representatives, these are shown in Table~\ref{table:rep_bitstrings}.
Once the representative states are found in cycle notation, we can construct the associated solitons (or states) per Eq.~\eqref{eq:cycle_soliton}.
This step is important since, e.g., for our particular circuit $(00101)$ and $(00011)$ belong to different equivalence classes, but the solitons $\leg \leg \leg \soliton \soliton$ and $\leg \leg \soliton \leg \soliton$ do not.
Additionally, Algorithm~\ref{alg:FKM-algorithm} can be readily used for qudits, as it appears in its original form for $d$-ary necklaces~\cite{ruskeyGeneratingNecklaces1992a}.

\begin{figure}
    \centering
    \includegraphics[width=0.66\columnwidth]{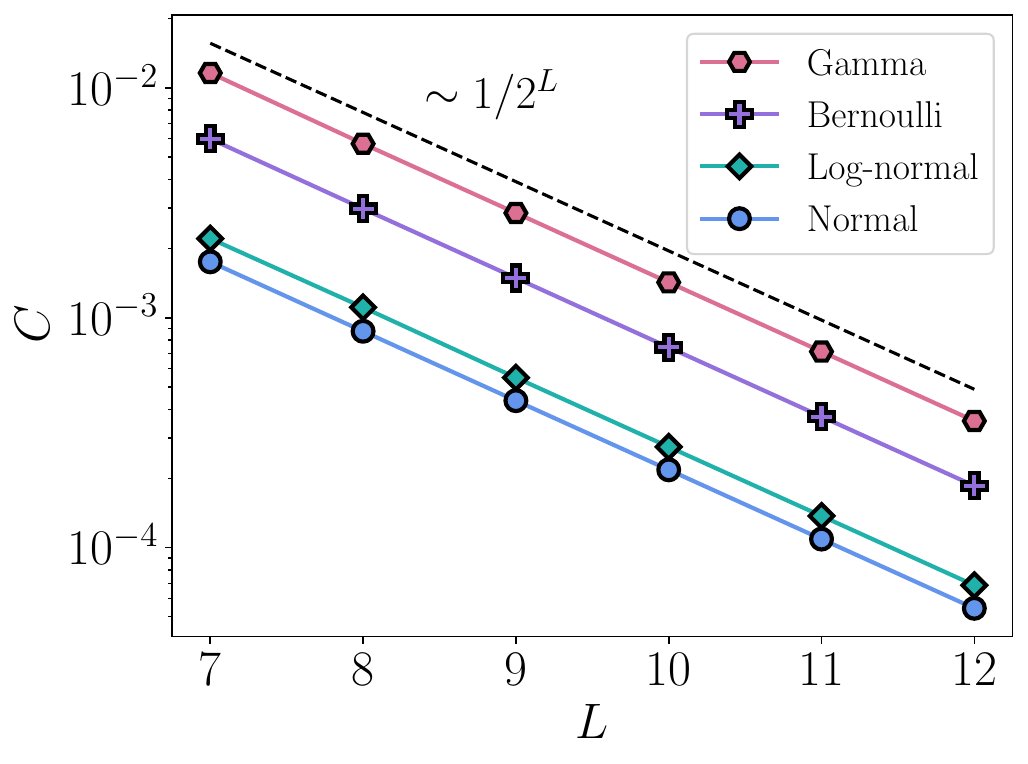}
    \caption{Crossing contributions of random Wigner matrix ensembles for different underlying distributions, according to the ETH definition from Eq.~\eqref{eq:crossing} for $t = 0$, i.e. the fourth moment of off-diagonal entries. The matrices have dimension $D = 2^L$.}
    \label{fig:RMT_crossing}
\end{figure}

\section{Wigner matrices}\label{app:rmt_crossing}

In this Appendix we numerically compare crossing diagrams in different random matrix distributions resulting in so-called Wigner matrices.
We construct arbitrary (symmetric) random matrices of dimension $N \times N$, where the elements $\tilde{A}_{ij}$ are sampled from four different classes of distributions: (i) from a normal distribution, (ii) a log-normal distribution, (iii) the Bernoulli distribution (with outcomes $\{0, 1\}$) and (iv) the gamma distribution. 
The parameters of the functions are chosen in a way that their mean and variance coincide. 
Afterwards, we normalize these matrices with $A_{ij} = \tilde{A}_{ij}/\sqrt{D}$ and compute the average of the crossing contributions for $100$ random realizations. 

The results are shown in Fig.~\ref{fig:RMT_crossing}. 
We can see that, regardless of underlying distribution, the suppression of crossing contributions is highly robust and scales with $1/D$ in all distributions.
We do however observe a small difference when it comes to the prefactor in these scaling relations; which suggests that in this case the particular choice of distribution does come into play.
Through the free probability machinery it can actually be shown that arbitrary Wigner matrices are also asymptotically free, although the combinatorics are more convoluted in some scenarios~\cite{speicherLectureNotesFree2019}.

Although there is a lack of the more involved structure present in (full) ETH, this basic random matrix model is consistent with our numerical results from Sec.~\ref{sec:nonint}.
Namely, it was observed that the particular shape of the distribution, be it Gaussian or log-normal, was inconsequential. 
Rather, the dimensionality of the Hilbert space, as dictated by the presence of symmetries, degeneracies and/or onset of chaos was the main ingredient in the observed scaling for these quantities.

Finally and somewhat tangentially, one might question why we observe these results even for individual stochastic realizations of the circuits, despite the fact that many of these functions and cumulants in FP are evaluated at the level of ensemble averages. The answer lies on the fact that many of these results, such as asymptotic freeness, hold a strong sense of almost sure convergence. 
In other words, almost all random realizations of the problem will agree with the averaged behavior~\cite{mingoFreeProbabilityRandom2017}. 




\bibliographystyle{apsrev4-2}

\bibliography{library}



\end{document}